\documentclass[useAMS,usenatbib]{mn2e}

\usepackage{times}
\usepackage{graphicx}

\title[Magnetic Field Amplification in a Relativistic Shock]{Magnetic Field Amplification and Saturation in Turbulence Behind a Relativistic Shock}
\author[Y. Mizuno et al.]{
Yosuke Mizuno$^{1}$\thanks{E-mail:mizuno@phys.nthu.edu.tw}, 
Martin Pohl$^{2, 3}$, Jacek Niemiec$^{4}$, Bing Zhang$^{5}$, Ken-Ichi Nishikawa$^{6}$,
\newauthor
 and Philip E. Hardee$^{7}$\\ 
$^{1}$Institute of Astronomy, National Tsing-Hua University, Hsinchu 30013, Taiwan, Republic of China\\
$^{2}$Institute of Physics and Astronomy, University of Potsdam, 14476 Potsdam, Germany\\
$^{3}$DESY, Platanenallee 6, 15738 Zeuthen, Germany\\
$^{4}$Institute of Nuclear Physics PAN, ul. Radzikowskiego 152, 31-342 Krak\'{o}w, Poland\\
$^{5}$Department of Physics and Astronomy, University of Nevada, Las Vegas, NV 89154, USA\\
$^{6}$Department of Physics, University of Alabama in Huntsville, Huntsville, AL 35805, USA\\
$^{7}$Department of Physics and Astronomy, The University of Alabama, Tuscaloosa, AL 35487, USA
}
\begin{document}

\date{Accepted 2014 January 27. Received 2013 December 2; in original form 2013 December 2}

\pagerange{\pageref{firstpage}--\pageref{lastpage}} \pubyear{2013}

\maketitle

\label{firstpage}

\begin{abstract}
We have investigated via two-dimensional relativistic MHD simulations the long-term evolution of turbulence created by a relativistic shock propagating through an inhomogeneous medium. In the postshock region, magnetic field is strongly amplified by turbulent motions triggered by preshock density inhomogeneities. Using a long-simulation box we have followed the magnetic-field amplification until it is fully developed and saturated. The turbulent velocity is sub-relativistic even for a strong shock. Magnetic-field amplification is controled by the turbulent motion and saturation occurs when the magnetic energy is comparable to the turbulent kinetic energy. Magnetic-field amplification and saturation depend on the initial strength and direction of the magnetic field in the preshock medium, and on the shock strength. If the initial magnetic field is perpendicular to the shock normal, the magnetic field is first compressed at the shock and then can be amplified by turbulent motion in the postshock region. Saturation occurs when the magnetic energy becomes comparable to the turbulent kinetic energy in the postshock region. If the initial magnetic field in the preshock medium is strong, the postshock region becomes turbulent but significant field amplification does not occur. If the magnetic energy after shock compression is larger than the turbulent kinetic energy in the postshock region, significant field amplification does not occur.
We discuss possible applications of our results to gamma-ray bursts and active galactic nuclei.
\end{abstract}


\section{Introduction}

Nonthermal emission is observed from many astrophysical sources harbouring relativistic shocks. In general, the composition of the plasma, the Lorentz factor of the shock, and the structure and strength of the preshock magnetic field are unknown. Radiation modeling of gamma-ray bursts (GRBs) suggests that the magnetic energy density in the emission region constitutes a substantial fraction 
$\epsilon_{B} \sim 10^{-3} - 10^{-1}$ of the internal energy density (e.g., Panaitescu \& Kumar 2002, Yost et al. 2003; Panaitescu 2005, Piran 2005; M\'{e}sz\'{a}ros 2006; Santana, Barniol Duran, \& Kumar 2013). However, such a high magnetization cannot be attained solely by a simple compressional amplification of the weak magnetic field pre-existing in the upstream plasma (Gruzinov 2001; Barniol Duran 2013).

Magnetic-field amplification beyond shock compression also seems necessary for emission modeling of young supernova remnants (SNRs), for which magnetic fields as strong as $\sim 1$ mG have recently been inferred from observations of the thin X-ray rims in several young SNRs (Bamba et al. 2003, 2005a, 2005b; Vink \& Laming 2003, but see also Pohl et al. 2005) along with rapid time variation of the synchrotron X-ray emission in RX J1713.7-3946 (Uchiyama et al. 2007, but see also Bykov et al. 2008).

Magnetic fields in GRB afterglow shocks can be generated through 
Weibel and filamentation instabilities (e.g., Medvedev \& Loeb 1999), as was demonstrated with Particle-in-cell (PIC) simulations of relativistic collisionless shocks (e.g., Nishikawa et al. 2005, 2009; Spitkovsky 2008). Recent studies have also proved that Weibel-type instabilities operate in sub-relativistic shocks (Kato \& Takabe 2008; Niemiec et al. 2012). It is a matter of debate whether magnetic fields thus generated will persist at sufficient strength over the entire emission region, that, for example, in GRBs is estimated to extend over some $10^6$ plasma skin depths downstream of the shock.

On larger scales, magnetic fields can be amplified through nonresonant cosmic-ray streaming instabilities in the precursor of nonrelativistic (e.g., Bell 2004; Niemiec et al. 2008; Riquelme \& Spitkovsky 2009, 2010; Stroman et al. 2009) and relativistic (e.g., Milosavljevi\'{c} \& Nakar 2006; Niemiec et al. 2010) shocks, which in the nonlinear phase induce density fluctuations. Upstream density fluctuations of any origin, e.g., those found in the wind zone of the GRB progenitor (e.g., Ramirez-Ruiz et al. 2005; Sironi \& Goodman 2007) or arising from cosmic-ray streaming instabilities (Stroman et al. 2009), can trigger a Richtmyer-Meshkov-type instability that leads to turbulent dynamo processes in the postshock region. 
Interaction of the shock front with such density fluctuations generates a significant vorticity at the shock. This turbulent plasma motion stretches and deforms magnetic field lines leading to field amplification (Sironi \& Goodman 2007; Goodman \& MacFadyen 2008; Palma et al. 2008).

The existence of relativistic turbulence in GRBs has been invoked to explain the observation of 
large variations in the prompt GRB $\gamma$-ray luminosity as well as intraburst variability in the afterglows. Narayan \& Kumar (2009) and Lazar et al. (2009) proposed a relativistic turbulence model instead of the well-known internal shock model to interpret the variable GRB light curves. However, the applicability of these models has recently been challenged by the results of MHD simulations performed by Inoue et al. (2011), who showed that relativistic turbulence decays much faster than the rate of  magnetic field amplification. Zhang \& Yan (2011) proposed a new GRB prompt-emission model in the highly magnetized regime, which invokes internal-collision-induced magnetic reconnection and turbulence. Within this model, turbulence could be sustained by continuous reconnection in the energy dissipation region. The short-time variability ``spikes'' in GRB lightcurves can be attributed to turbulent reconnection in the magnetic dissipation region, while the long-time variability stems from the activity of the central engine.

The blazar zone -- the innermost part of the relativistic jets in Active Galactic Nuclei (AGN) -- is probed through multi-wavelength observations. Aharonian et al. (2003) and Krawczynski et al. (2004) reported correlated X-ray/TeV $\gamma$-ray flares with timescales from 15 minutes (for Mrk 421) to a few hours (for Mrk 501 and 1ES 1959+650). In the TeV band alone, flux doubling has been observed on timescales down to 2 minutes \citep{2007ApJ...664L..71A, Alb07,2013ApJ...762...92A}. Huge Doppler factors ${\rm D}\ga 50$ appear required to provide $\gamma\gamma$ opacities $\tau \la 1$ and permit emission regions larger than the Schwarzschild radius of the central black holes. A scenario of fast-moving ``needles" within a slower jet or of a ``jet-within-a-jet" (Levinson 2007; Begelman et al. 2008; Ghisellini \& Tavecchio 2008; Giannios et al. 2009) has been invoked to explain the fast variability of blazars. The short-term fluctuations can be also understood as a consequence of a turbulent ambient-jet-plasma that passes through shocks in the jet flow (Marscher \& Jorstad 2010; Marscher et al.\ 1992).

The potential importance of turbulence to magnetic field amplification and variability led Giacalone \& Jokipii (2007) and Guo et al. (2012) to perform 2D non-relativistic MHD shock simulations involving upstream density and magnetic field fluctuations with a Kolmogorov power spectrum. Two- and three-dimensional MHD simulations of non-relativistic shocks propagating in a cloudy inhomogeneous interstellar medium (ISM) have been also performed by \citet{Ino09, Ino12}. The simulations indicated strong magnetic-field amplification in the postshock medium. The peak magnetic-field strength was found to be more than a hundred times larger than the preshock field strength. Similar results have been also obtained in recent hybrid (kinetic ions and fluid electrons) simulations by Caprioli \& Spitkovsky (2013).
Fraschetti (2013) has investigated magnetic field amplification by turbulence generated downstream of a two-dimensional rippled hydromagnetic shock analytically in non-relativistic regime. These results strengthen the case for turbulence being an important contributor to  magnetic-field amplification and emission variability.

In an earlier paper (Mizuno et al. 2011b), we demonstrated that the magnetic field is amplified by the turbulence that develops in the post-shock region behind a relativistic shock propagating through an inhomogeneous medium. Inoue et al. (2011) performed 3D relativistic MHD simulations of a propagating relativistic shock and obtained results similar to ours. However, the growth of the magnetic field had not saturated in the relatively short time covered by this previous work.

In this paper, we continue our investigation and present results from two-dimensional (2.5D) relativistic MHD simulations using a much longer grid. This longer grid permits us to investigate the long-term evolution of turbulence and the saturation of magnetic-field amplification.  We also extend our investigation to non-relativistic in addition to relativistic shock speeds and to a range of magnetizations of the upstream medium.

This paper is organized as follows: We describe the numerical method and setup used for our simulations in \S 2, present our results in \S 3, and discuss their astrophysical implications in \S 4.

\section{Numerical Method and Setup}

We solve the 3D RMHD equations for a mildly relativistic shock propagating in an inhomogeneous medium in two-dimensional Cartesian geometry ($x-y$ plane), but follow all three components of the velocity and magnetic field vectors (so-called 2.5D or 2D3V model) using the 3D GRMHD code ``RAISHIN"  (Mizuno et al. 2006, 2011a).
For the simulations described here, we have introduced a fifth-order weighted essentially non-oscillatory (WENO) scheme.  

The setup of the code for shock simulations was outlined in Mizuno et al. (2011b). However, in these simulations we use a computational box that is 4 times longer, namely $(x,y)=(8L,L)$. The numerical resolution is as before with $N/L=256$. At $x=x_{max}$, the fluid which is initially moving with velocity $v_{x}=v_{0}$ in the positive $x$-direction is stopped by setting $v_x=0$ and thermalized
\footnote{This condition for the boundary is different from a typical reflecting boundary, at which $v_{x} = -v_{x}$ (e.g., Spitkovsky 2008). The advantage of our approach is that the conversion of kinetic to thermal energy at the boundary closely mimics the fluid behaviour behind a shock and in front of the contact discontinuity.  Additionally, the created shock propagates slower and we can follow it for a longer time.}.
As the pressure increases at $x=x_{max}$, a shock forms and propagates in the $-x$-direction. 
The downstream plasma velocity is thus zero on average.  
To produce different shock strengths, we choose three different flow speeds, $v_{0}=0.2c$, $0.5c$ and $0.9c$, where $c$ is the speed of light. 

As in our previous work (Mizuno et al. 2011b), simulations are initialized with an inhomogeneous plasma with mean rest-mass density $\rho_{0}=1$ containing fluctuations $\delta \rho$ established across the entire simulation domain. Following Giacalone \& Jokipii (1999, 2007),  density fluctuations are created by superposing 50 discrete wave modes with wavelengths between $\lambda_{min}=0.025L$ and $\lambda_{max}=0.5L$. The wave amplitudes are chosen to mimic a two-dimensional Kolmogorov-like power-law spectrum given by
\begin{equation}
P_{k} \propto {1 \over 1 + (kL)^{8/3}},
\end{equation}
where $L$ is the turbulence coherence length.
The fluctuation variance is $\sqrt{\langle\delta \rho^{2}\rangle} = 0.012 \rho_{0}$.
Note, that in contrast to the method used by Giacalone \& Jokipii (2007), our initial preshock turbulence does not include any fluctuating magnetic field (see also Mizuno et al. 2011b).

The gas pressure of preshock medium is a constant with $p=0.001\rho_{0}c^{2}$, which is an order of magnitude lower than the value used in Mizuno et al. (2011b).
The shock waves that form may thus have a larger sonic Mach number. An equation of state (EoS) relates the enthalpy $h$ to the gas-pressure and density, and here we use the so-called TM EoS proposed by Mignone et al.(2005): 
\begin{equation}
h={5 \over 2} \Theta + \sqrt{{9 \over 4 } \Theta^2 + 1},
\end{equation}
where $\Theta \equiv p/(\rho c^2)$. The TM EoS is a simple algebraic function of $\Theta$ and a good approximation to Synge's EoS (Synge 1971) that describes  single-component perfect gases in the relativistic regime. 
The TM EoS corresponds to a lower bound of Taub's fundamental inequality (Taub 1948), i.e., $(h-\Theta)(h-4\Theta)=1$. The TM EoS reproduces the correct asymptotic values for the equivalent adiabatic index $\Gamma_{eq} =  (h-1)/( h-1-\Theta)$, i.e. $\Gamma_{eq} \to 5/3$ for non-relativistic temperatures and $\Gamma_{eq} \to 4/3 $ in the ultra-relativistic limit.

In our simulations the pre-shock plasma carries a constant mean magnetic field. To investigate the effect of the initial magnetic field strength, we choose three different magnetizations in the pre-shock medium, $\sigma \equiv b^2 / \rho c^2 = 0.0001$, $0.001$, and $0.01$. Here $\rho$ is the density and $b$ is the magnetic
field in the comoving (pre-shock medium) frame. Note that $b^2 = \mathbf{B}^{2}/\gamma^{2} + (\mathbf{v} \cdot \mathbf{B})^{2}$ where $\mathbf{B}$ is the magnetic field seen in the simulation frame (Komissarov 1997; Del Zanna et al. 2007). We also consider two different magnetic-field orientations with respect to the shock normal, a parallel (to the shock normal, $B_{x}$, $\theta_{Bn}=0\degr$) and a perpendicular ($B_{y}$, $\theta_{Bn}=90\degr$) field configuration.  The sound speed, $c_{s}/c=[\Theta (5h-8\Theta)/3h(h-\Theta)]^{1/2}$, and the Alfv\'{e}n speed, $v_{A}/c=[b^2/(\rho h + b^2)]^{1/2}$, in the different simulations are calculated using the mean plasma density ($\rho_{0}$) measured in the comoving (pre-shock medium) frame , and are listed in Table 1 along with the flow speeds $v_{0}$.

\begin{table}
  \caption{Simulation Parameters} \label{table1}
     \begin{tabular}{lcccccc}
     \hline
     \hline
     Case & $v_{0}$ & $\sigma$ & $B_0$ & $\theta_{Bn}$ & $c'_{s}/c$ & $v'_A/c$ \\
     \hline
     A1 & 0.5 & 0.0001 & 0.01 & 0 & 0.0408 & 0.0115\\
     A2 & 0.5 & 0.001  &  0.032 & 0 &  0.0408 & 0.0365\\
     A3 & 0.5 & 0.01  &  0.1 & 0 &  0.0408 & 0.114\\
     B1 & 0.5 & 0.0001 &  0.0115 & 90 &  0.0408 & 0.01\\
     B2 & 0.5 & 0.001 &  0.0365 & 90 &  0.0408 & 0.0315\\
     B3 & 0.5 & 0.01 &  0.115 & 90 &  0.0408 & 0.1\\ 
     C1 & 0.9 & 0.0001 & 0.01 & 0 &  0.0408 & 0.01\\
     C2 & 0.9 & 0.001  &  0.032 & 0 &  0.0408 & 0.0315\\
     C3 & 0.9 & 0.01  &  0.1 & 0 &  0.0408 & 0.1\\
     D1 & 0.9 & 0.0001 &  0.023 & 90 &  0.0408 & 0.01\\
     D2 & 0.9 & 0.001 &  0.0725 & 90 &  0.0408 & 0.0315\\
     D3 & 0.9 & 0.01 &  0.23 & 90 &  0.0408 & 0.1\\
     E1 & 0.2 & 0.0001 & 0.01 & 0 &  0.0408 & 0.0102\\
     E2 & 0.2 & 0.001  &  0.032 & 0 &  0.0408 & 0.0322\\
     E3 & 0.2 & 0.01  &  0.1 & 0 &  0.0408 & 0.101\\
     F1 & 0.2 & 0.0001 &  0.01 & 90 &  0.0408 & 0.01\\
     F2 & 0.2 & 0.001 &  0.032 & 90 &  0.0408 & 0.0316\\
     F3 & 0.2 & 0.01 &  0.1 & 90 &  0.0408 & 0.1\\     
      \hline     
     \end{tabular}
\end{table}

\section{Results}

\subsection{Dependence on the Initial Magnetic Field Strength}

In this section we describe the main characteristics of the turbulence generated when a shock propagates into an inhomogeneous medium, and the dependence of the properties of the system on the preshock plasma magnetization. As a representative example of this dependence (studied here for all the cases considered, see Table 1), we choose the case with the mildly relativistic flow speed of $v_{0}=0.5c$ and a parallel configuration for the mean preshock magnetic field with respect to the shock normal, $\theta_{Bn}=0\degr$ (cases A1-A3). In the following, we refer to the three different plasma magnetizations studied in this paper, $\sigma=0.0001$, $0.001$, and $0.01$, as the {\emph{low-}}, {\emph{medium-}}, and {\emph{high}}-$\sigma$ cases, respectively.

\subsubsection{Global Structure}

Figures 1 and 2 show 2-D images of the density (Fig.\ 1) and the total magnetic-field strength (Fig.\ 2)
\begin{figure*}
\begin{center}
\includegraphics[width=14cm]{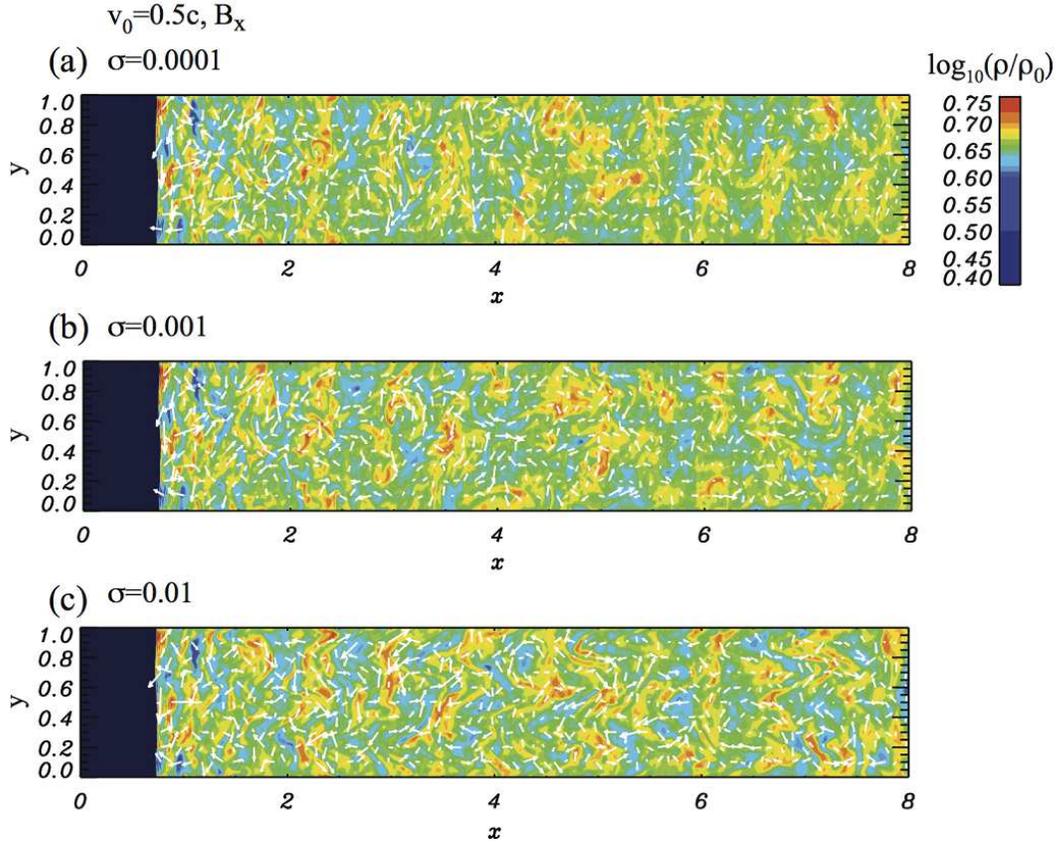}
\caption{Two-dimensional images of density at $t_{s}=42$ for magnetization parameters $\sigma=0.0001$ ({\it upper panel}), $0.001$ ({\it middle}), and $0.01$ ({\it lower panel}) for case A, with magnetic field parallel to the shock normal and $v_{0}=0.5c$. White arrows indicate the flow direction in the postshock region.
\label{f1}}
\end{center}
\end{figure*}
at $t_{s}=42$, where $t_{s}$ is in units of $L/c$ with $c=1$, for the three magnetization parameters studied in case A. 
\begin{figure*}
\begin{center}
\includegraphics[width=14cm]{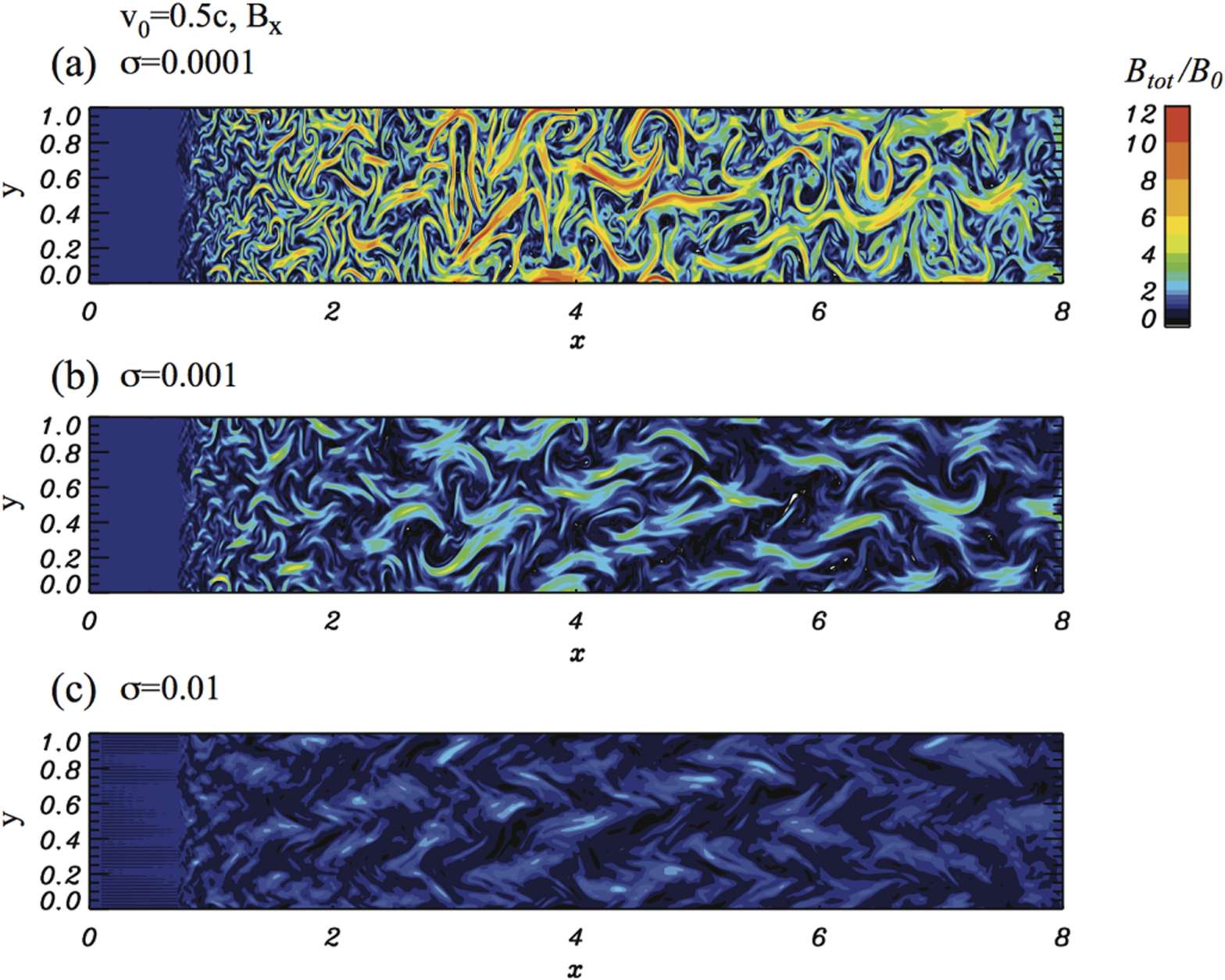}
\end{center}
\caption{Two-dimensional images of the total magnetic field normalized to the initial magnetic field strength, $B_{0}$, at $t_{s}=42.0$ for magnetization parameters $\sigma=0.0001$ ({\it upper panel}), $0.001$ ({\it middle}), and $0.01$ ({\it lower panel}). As in Fig.~\ref{f1} the magnetic field is parallel to the shock normal and $v_{0}=0.5c$ (case A).
\label{f2}}
\end{figure*}
As described in Mizuno et al. (2011b), the shock front develops ripples when the inhomogeneous-density preshock plasma encounters the shock. In all cases these ripples lead to strong, random transverse flow behind the shock, thus introducing rotation and vorticity in the postshock region through a process similar to the Richtmyer-Meshkov instability (e.g., Brouillette 2002; Sano et al. 2012; Inoue 2012). The turbulent plasma motions produce the velocity shears along magnetic field lines that lead to the magnetic-field amplification. Note that because our simulations start from pre-existing finite-amplitude density fluctuations in the preshock medium, the flow pattern in the postshock region is initially highly nonlinear and comparison with linear Richtmyer-Meshkov instability analysis is not useful. 

In the low-$\sigma$ case ($\sigma$=0.0001, run A1), the preshock magnetic field energy density is much less than the postshock turbulent energy density. Thus, the turbulent velocity field can easily stretch and deform the frozen-in magnetic field, resulting in field-amplification. Near the shock front, the vorticity scale size is small, but farther downstream the scale of the vortices increases through an inverse cascade of turbulent eddies, and the magnetic field is strongly amplified. The turbulent density structure is nearly isotropic because the magnetic field is weak. The amplified magnetic field develops a filamentary structure. In the region far behind the shock front at $x \sim 6-8$, the magnetic-field strength decreases relative to that at $x \sim 3-5$, where the highest amplification is observed. This indicates that magnetic-field amplification via turbulent motion saturates and then tends to decay farther behind the shock (see Fig. 4 for the time evolution of the volume-averaged magnetic field).

In the medium-$\sigma$ case ($\sigma=0.001$, run A2), the preexisting magnetic field energy density is still less than the postshock turbulent energy density. As in the  low-$\sigma$ case, the downstream magnetic field is amplified in the turbulent velocity field. The magnetic field is structured in thicker filaments than are seen in the low-$\sigma$ case.
The magnetic filaments are aligned along the initial magnetic-field direction, $B_{x}$, after saturation, most likely because magnetic-field tension resists motion perpendicular to the mean-field direction.

In the high-$\sigma$ case ($\sigma=0.01$, run A3), magnetic-field amplification through the turbulent dynamo process is not efficient, even though a turbulent velocity field develops in the postshock region. The magnetic field is amplified only by a factor of about 2 relative to the initial magnetic field. Magnetic filamentary structures seen in the lower-$\sigma$ cases are significantly suppressed at this higher $\sigma$ value.

Figure 3 shows 1-D cuts along the $x$-axis at $y/L=0.5$ and $t_{s}=42$ of the density, the in-plane transverse velocity $v_{y}$, and the total magnetic field strength for parallel magnetic field with magnetization parameters $\sigma=0.0001$ (black solid), $0.001$ (red dotted), and $0.01$ (blue dashed) and the mildly relativistic flow velocity $v_{0}=0.5c$ (runs A1-A3).
The shock front is located at $x\simeq 0.8L$. The left and right sides of the shock front are upstream and downstream regions, respectively. The measured shock propagation speed is about $v_{sh} \simeq 0.17c$ in the contact discontinuity frame. Analytic calculations using an ideal gas EoS (see the Appendix in Mizuno et al. 2011b) give a shock velocity in the contact-discontinuity frame of $\sim 0.18c$ for $\Gamma=5/3$ and $\sim 0.09c$ for $\Gamma=4/3$. The measured shock velocity is thus in good agreement with that expected for $\Gamma=5/3$. Conventionally, the shock velocity and Mach number are given in the upstream rest frame.
We parametrize the shock strength by the relativistic sonic Mach number 
\begin{equation}
M_{s} \equiv \gamma_{sh}^\prime v_{sh}^\prime/\gamma_{s}^\prime c_{s}^\prime,
\label{mach}
\end{equation}
where $\gamma_{s}^\prime \equiv (1-c_{s}^{\prime \,2}/c^2)^{-1/2}$ is the Lorentz factor associated with the sound speed and $v_{sh}^\prime$ and $\gamma_{sh}^\prime$ are the shock speed and the shock Lorentz factor measured in the upstream rest frame. 
A trivial Lorentz transformation converts the shock propagation speed measured in the simulation to the standard upstream-frame shock speed,
\begin{equation}
v_{sh}^\prime =\frac{v_{sh}+v_{0}}{1+v_{sh} v_{0}/c^2}.
\label{shockspeed}
\end{equation}
The shock propagation speed of $v_{sh}=0.17c$ obtained from the simulations thus corresponds to $v_{sh}^\prime \simeq 0.82c$ in the upstream flow frame. This leads to $M_{s}\simeq 19$ for the sound speed in the preshock region where $c_{s}^\prime\simeq 0.04c$.
\begin{figure}
\begin{center}
\includegraphics[width=8.5cm]{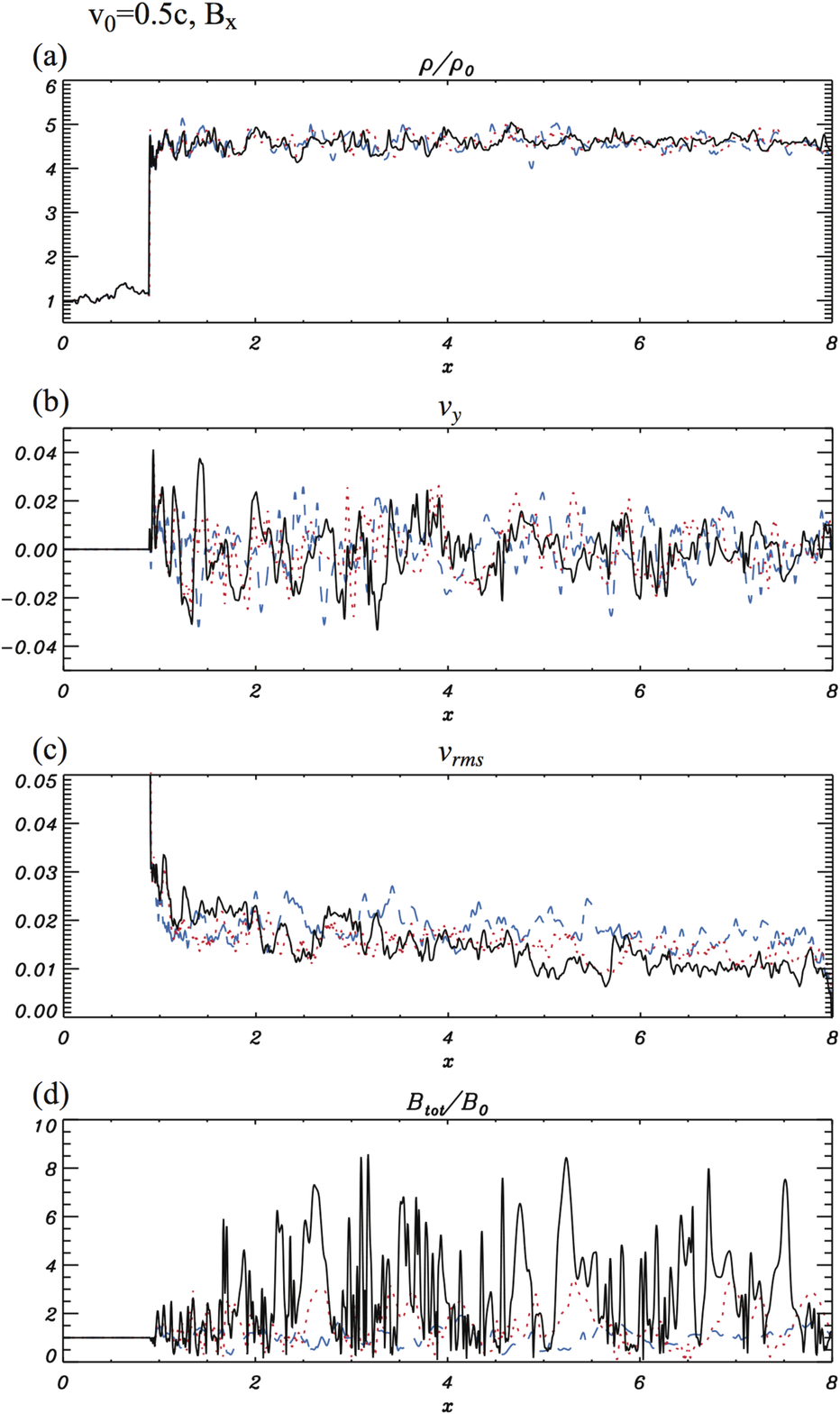}
\end{center}
\caption{One-dimensional cuts  along the $x$-direction of ({\it a}) the normalized density ($\rho / \rho_{0}$), ({\it b}) the transverse velocity ($v_{y}$), ({\it c}) the root-mean-square velocity averaged in the $y-$ direction ($v_{rms}$) and ({\it d}) the total normalized magnetic field strength ($B_{tot}/B_{0}$), at $y/L=0.5$ and $t_{s}=42.0$ for parallel magnetic field cases with $\sigma=0.0001$ (black solid lines), $\sigma=0.001$ (red dotted lines), and $\sigma=0.01$ (blue dashed lines) and mildly relativistic flow velocity $v_{0}=0.5c$ (cases A1-A3). The shock front is located at $x/L\simeq 0.8$. The left and the right sides of the shock front are upstream and downstream regions, respectively.
\label{f3}}
\end{figure}

In all cases, the density jumps by about a factor of 4, which is close to the strong-shock limit in the Newtonian approach. The transverse-velocity is strongly fluctuating. The maximum transverse velocity is about $0.04c$, and in all cases the average root mean square turbulent velocity of $\sim 0.02c$ is subsonic in the postshock region ($\langle c_{s}\rangle \simeq 0.35c$). The total-magnetic-field also shows strong variation. The magnetic field is not compressed at the shock because the direction of the initial magnetic field is parallel to the shock normal. In the low-$\sigma$ case ($\sigma=0.0001$), the local magnetic field reaches nearly 8 times the amplitude of the initial field.  When the initial magnetic field is larger, magnetic-field amplification is reduced. In the medium-$\sigma$ case ($\sigma=0.001$), the maximum amplitude of the amplified magnetic field is $\langle B_{\rm tot}\rangle/B_0\sim 4$, whereas in the high-$\sigma$ case ($\sigma=0.01$), the local magnetic field in the postshock region reaches only 2 times the initial magnetic field strength. The Alfv\'{e}n velocity in the postshock region depends on the initial mean magnetic field strength. In the low-$\sigma$ case, the Alfv\'{e}n velocity in the postshock region  fluctuates strongly and the average Alfv\'{e}n velocity is $\langle v_{A}\rangle \simeq 0.01c$. 
Turbulence is super-Alfv\'{e}nic in most of the postshock region in the low-$\sigma$ case. This result is consistent with earlier non-relativistic studies (e.g., Giacalone \& Jokipii 2007; Inoue et al. 2009; Guo et al. 2012). 
When the mean magnetic field is larger, the Alfv\'{e}n velocity in the postshock region is larger. In the middle- and high-$\sigma$ cases, the average Alfv\'{e}n velocities in the postshock region are $\langle v_{A}\rangle \simeq 0.03c$ and $0.06c$ respectively. Postshock turbulence in the medium- and high-$\sigma$ cases is sub-Alfv\'{e}nic.

\subsubsection{Magnetic Field Amplification \& Saturation}

Previous results showed that magnetic-field amplification via the  turbulent dynamo process depends on the initial magnetic-field strength. Figure 4 shows the time evolution 
of the volume-averaged total magnetic field (Fig. 4a) and the maximum total magnetic-field strength (Fig. 4b) in the postshock region for cases A1-A3. We continuously check the shock position for each $y$-coordinate and average the absolute magnetic-field strength from the y-dependent shock position to $x=x_{max}$. The region over which the average is taken thus explicitly depends on time and the case in study.
All cases are normalized by the initial magnetic-field strength, $B_0$, which varies with the assumed $\sigma$.
\begin{figure}
\begin{center}
\includegraphics[width=8.5cm]{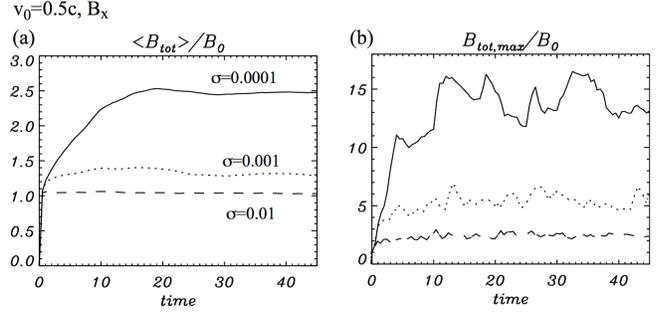}
\end{center}
\caption{Time evolution of ({\it a}) the volume-averaged total magnetic field and ({\it b}) the maximum total magnetic-field strength in the postshock region normalized by $B_0$ for the case of 
mildly relativistic flow velocity ($v_{0}=0.5c$) and parallel mean magnetic field configuration (runs A1-A3). Different lines are for different initial magnetic field strength: $\sigma=0.0001$ (solid lines), $\sigma=0.001$ (dotted lines), and $\sigma=0.01$ (dashed lines). 
\label{f4}}
\end{figure}

In the low-$\sigma$ case ($\sigma=0.0001$), the average postshock magnetic field gradually increases with time, saturates at $t_{s} \sim $ 20, and then decreases slightly to assume a constant value for $t_{s} \ga 30$. The postshock magnetic field is amplified by about a factor of 2.5 at saturation. The peak field strength is much larger than the mean field amplitude and about $16$ times larger than the initial magnetic-field strength. In the medium-$\sigma$ case ($\sigma=0.001$), the mean postshock field also gradually increases with time, but saturates at $t_{s} \sim 10$, sooner than in the low-$\sigma$ case, and is amplified by only a factor of 1.3. The peak field strength is about $7$ times larger than the initial magnetic-field strength. In the high-$\sigma$ case ($\sigma=0.01$), the mean magnetic field does not become stronger with time, thus the average magnetic field is not amplified in this case. The peak field in the postshock region is about 2 times larger than the initial magnetic field, suggesting some localized field amplification. These results show that the efficiency of magnetic-field amplification declines as the magnetization ($\sigma$) increases.

\subsubsection{The Kinetic to Magnetic Energy Density Ratio}

Figure 5 shows  2-D images of the ratio of the kinetic to the magnetic energy at $t_{s}=42$ for cases A1-A3.
The kinetic and magnetic energy densities are defined as $E_{kin}= (\gamma-1)\rho c^2$ and $E_{mag}=B^2 + [v^2 B^2 - (\vec{v} \cdot \vec{B})^2] /2$, respectively. In the low-$\sigma$ case, the kinetic energy dominates near the shock front ($x/L \la 2$).  In an intermediate region farther downstream from the shock at $2 \la x/L \la 5$, the magnetic energy becomes dominant as the magnetic field is amplified. Far behind the shock front ($x/L \ga 5$), where the field amplitude has saturated, the magnetic energy dominates in most locations.

Saturation occurs when the magnetic energy density becomes comparable to the turbulent kinetic energy density, in agreement with previous MHD studies (e.g., Schekochihin \& Cowley 2007; Cho et al. 2009; Zhang et al. 2009; Inoue et al. 2011). For cases A1-A3 with mildly relativistic flow speed, $v_{0}=0.5c$, and parallel magnetic field configuration, the root mean square turbulent velocity is $v_{rms} \equiv \sqrt{\langle v^2_{turb}\rangle} \simeq 0.02c$ (see Fig. 3c). The average kinetic energy density can be estimated from $\langle E_{kin} \rangle \sim \langle \rho_{d} \rangle v^2_{rms}/2 \sim 8 \times 10^{-4}$, where $\langle \rho_{d} \rangle \simeq 4$ is the average density in the postshock (downstream) region. Here we use the Newtonian approximation because the turbulence is not relativistic. If the magnetic field is amplified to the limit, i.e., the magnetic energy density becomes comparable to the turbulent kinetic energy density, then the estimated magnetic field strength at saturation in the laboratory (contact-discontinuity) frame is $\langle B_{\rm sat, est} \rangle \sim \sqrt{2 E_{kin}} \simeq 0.04$.  In the low-$\sigma$ case A1 the average magnetic field strength in the saturation region is $\langle B_{\rm sat, sim} \rangle \sim 0.035$. This simulation result is in good agreement with the estimate.
A similar result is found for the medium-$\sigma$ case A2, and at saturation the magnetic energy density is comparable to the turbulent kinetic energy density in this case as well.
In the high-$\sigma$ case A3, the estimated magnetic-field saturation level is lower than the initial field strength, even for the maximum turbulent velocity. Therefore, significant field amplification does not occur in this case.
\begin{figure*}
\begin{center}
\includegraphics[width=14cm]{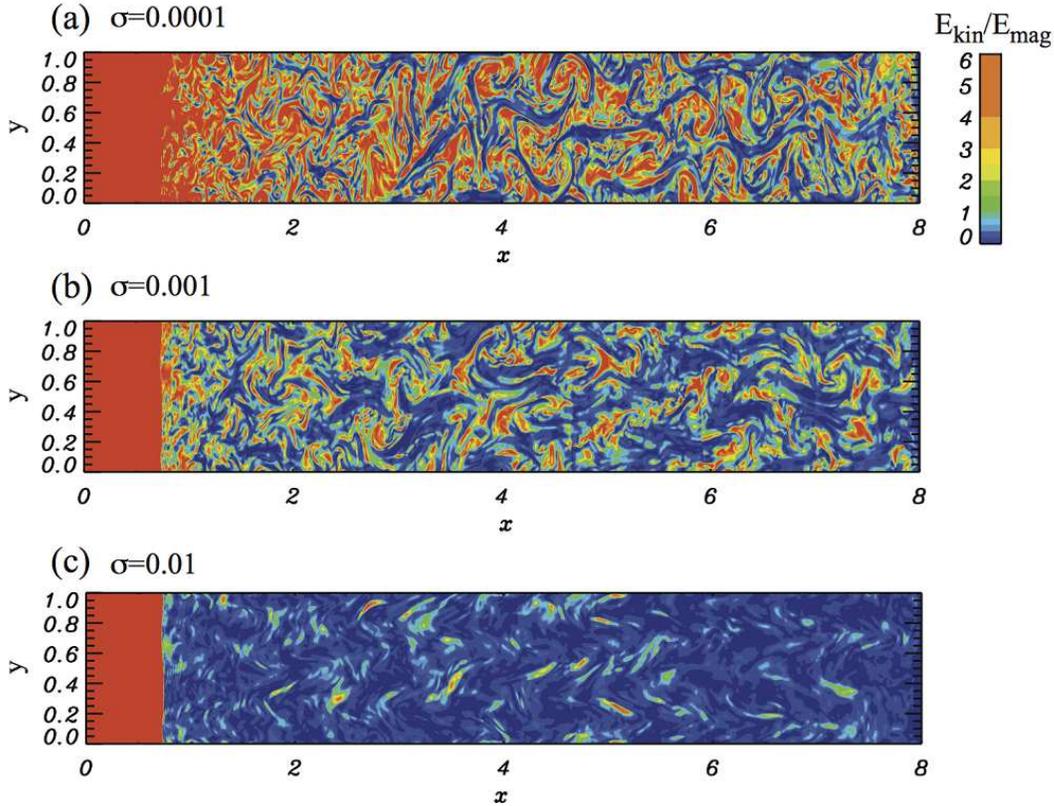}
\end{center}
\caption{Two-dimensional images of the kinetic to magnetic energy density ratio at $t_{s}=42.0$ for magnetization parameters, $\sigma=0.0001$ ({\it upper}), $0.001$ ({\it middle}), and $0.01$ ({\it lower}) with mean magnetic field parallel to the shock normal and $v_{0}=0.5c$ (cases A1-A3). 
\label{f5}}
\end{figure*}

subsubsection{Turbulent Magnetic and Kinetic Energy Power Spectra}

The statistical properties of turbulent fluctuations in the postshock region can be determined from their power spectra. Figure 6 shows spherically-integrated  kinetic and magnetic energy spectra for cases A1-A3.
As found by Mizuno et al. (2011b), the kinetic-energy spectra are only slightly flatter than Kolmogorov, i.e.,  $E_{kin}(k) \propto k^{-(5/3)-(D-1)}$ with $D=2$ in two-dimensional systems.
The kinetic-energy power spectra do not change significantly with time and are almost the same for all $\sigma$-cases. A Kolmogorov-like kinetic energy power spectrum seems to be an inherent property of the postshock turbulence produced by the interaction of the shock front with the upstream density inhomogeneities, since this power spectrum is observed both in studies that assume a Kolmogorov (e.g., Mizuno et al. 2011b) and also a non-Kolmogorov (e.g., Inoue et al. 2009) power spectrum in the upstream fluctuations. 

\begin{figure*}
\begin{center}
\includegraphics[width=12cm]{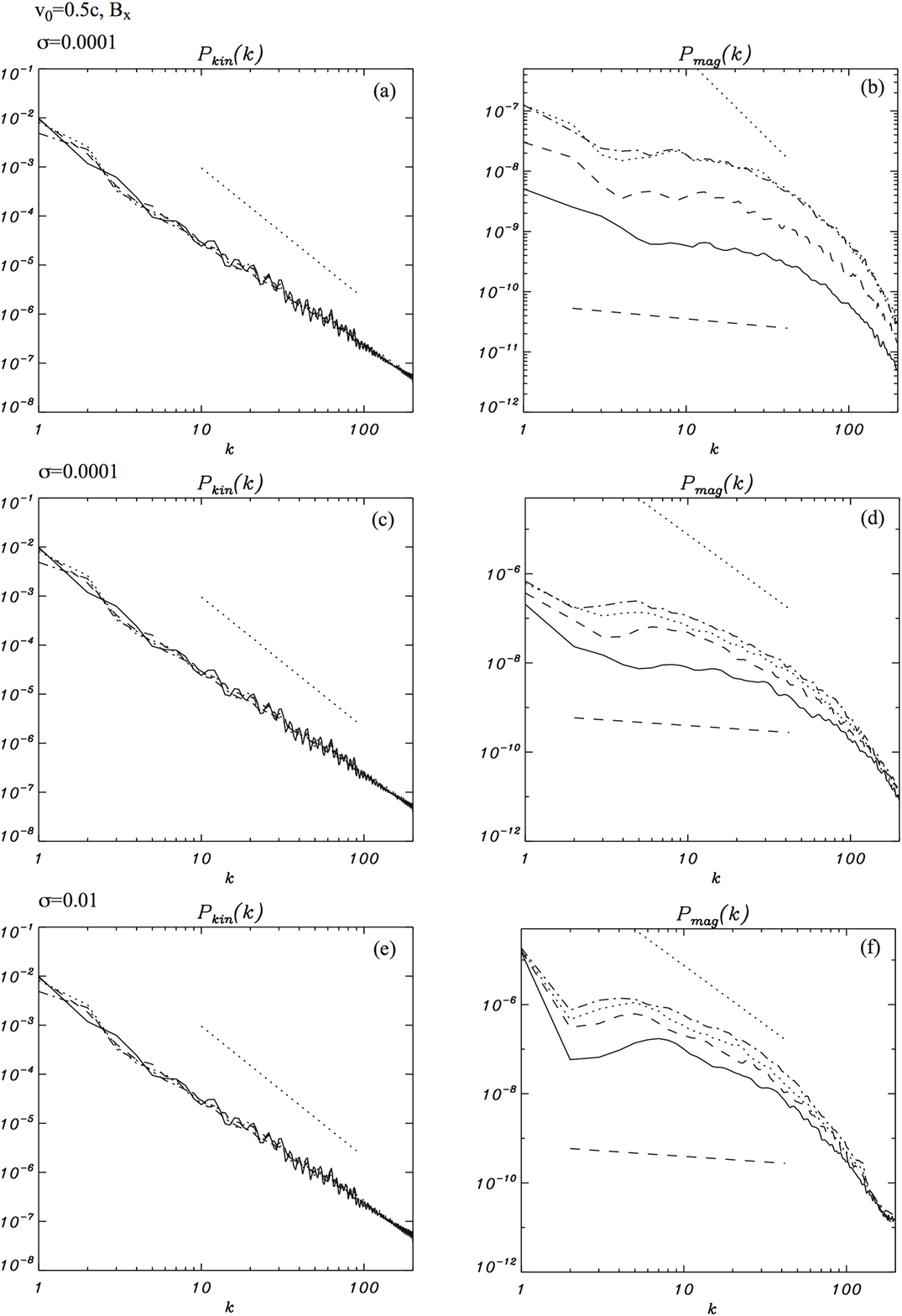}
\end{center}
\caption{Spherically-integrated power spectra of ({\it a, c, e}) the kinetic energy, ({\it b, d, f}) the magnetic energy in the postshock region for parallel magnetic field with $\sigma=0.0001$ (upper panel), $\sigma=0.001$ (middle), and $\sigma=0.01$ (lower panel) and mildly relativistic flow $v_{0}=0.5c$ (cases A1-A3). Different lines denote the spectra generated at different times: $t_{s}=3$ ({\it solid}), $8$ ({\it dashed}), $13$ ({\it dotted}), and 17 ({\it dash-dotted}).  A short dotted line, representing the 2D Kolmogorov power law $E(k) \sim k^{-8/3}$ and a short dashed line following $k^{-1/4}$ are shown for comparison in all the panels. 
\label{f6}}
\end{figure*}

In all cases, the magnetic energy power spectrum amplitude rapidly increases at early simulation times and the shape remains almost constant at later times, implying that magnetic-field amplification has reached saturation. The largest enhancement in the magnetic energy power spectrum occurs for the low-$\sigma$ case and  this reflects a larger amplification than for the higher $\sigma$-cases.
Consistent with Mizuno et al. (2011b), the magnetic energy spectra at large scales ($k\la 50$) are almost flat and strongly deviate from the Kolmogorov spectrum. Such spectra are typical of the small-scale dynamo process (Kazantsev 1968). Flat magnetic-energy spectra are produced in turbulent-dynamo simulations (e.g., Schekochihin et al. 2004; Brandenburg \& Subramanian 2005). The same properties are also observed in simulations of driven super-Alfv\'{e}nic turbulence (e.g., Cho \& Lazarian 2003) and in relativistic MHD turbulence simulations (e.g., Zhang et al. 2009; Inoue et al. 2011). Note that the magnetic energy spectrum in the low- and medium-$\sigma$ cases is flat over a broader region than the high-$\sigma$ case. This is again an indication of larger magnetic-field amplification for lower magnetizations.

\subsection{Dependence on Initial Magnetic Field Direction}

Figure 7 shows  2-D images of the total magnetic-field strength for the three different magnetization parameters and a mildly relativistic flow velocity $v_0 = 0.5c$ but now for mean magnetic field perpendicular to the shock normal, $\theta_{Bn}=90\degr$ (cases B1-B3). The low- and medium-$\sigma$ runs are shown at $t_{s} = 42$, whereas the high-$\sigma$ case is shown at $t_{s}=38$.   
\begin{figure*}
\begin{center}
\includegraphics[width=14cm]{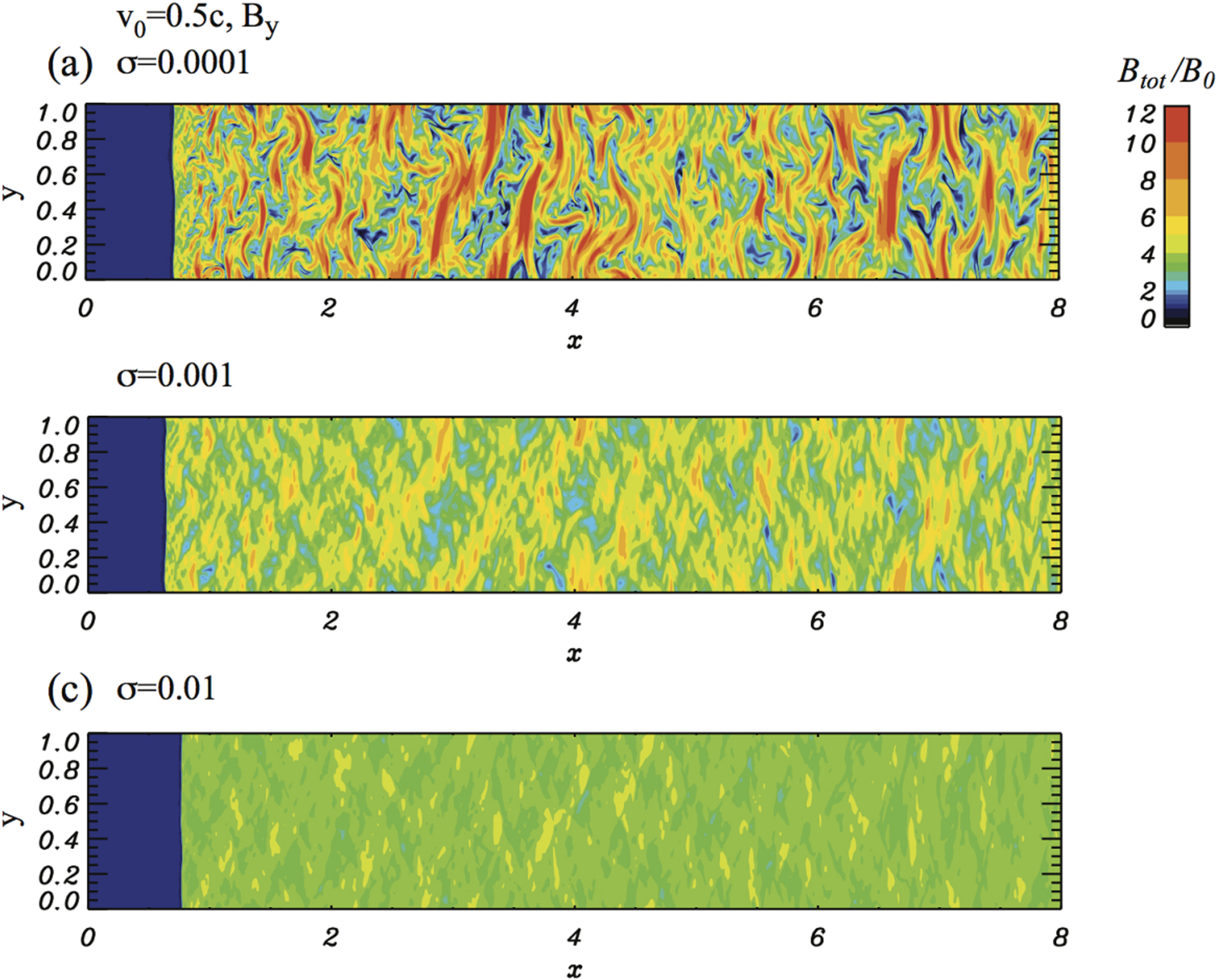}
\end{center}
\caption{The same as Fig. 2 but for magnetic field perpendicular to the shock propagation direction (cases B1-B3) at simulation time $t_{s} = 42$ for $\sigma=0.0001$ ({\it upper panel}) and $\sigma=0.001$ ({\it middle panel}), and $t_{s}=38$ for $\sigma=0.01$ ({\it lower panel}).
\label{f7}}
\end{figure*}
When the initial field direction is perpendicular to the shock normal, the magnetic field in the postshock region is shock-compressed by about a factor of 4 (see the 1D plot in Fig. 8). In all cases,  turbulence develops  in the postshock region as the relativistic shock passes through the inhomogeneous preshock medium. In the low-$\sigma$ case, after shock-compression, the magnetic field is still weak enough to be twisted up by the turbulent motion. Therefore, efficient magnetic-field amplification via the turbulent dynamo occurs and the magnetic field develops filamentary structure similar to that seen for the case with parallel mean magnetic field. However, the filaments are thicker in this case and aligned in the $y$-direction (the initial magnetic-field direction). Turbulent structure is not isotropic as it was in the low-$\sigma$ case with parallel magnetic field. This difference is a result of the shock compression of the magnetic field that does not occur for a parallel field configuration, and the stronger postshock field's influence on the turbulent motion. 
In the medium-$\sigma$ case, after the initial shock compression, the magnetic field is only moderately amplified in the postshock region. It appears that the compressed magnetic field is too strong to allow a significant deformation of the field lines by the turbulent velocity field. For the same reason, in the high-$\sigma$ case basically only shock compression is observed.

One-dimensional cuts along the $x$-axis at $y/L = 0.5$ and $t_{s} = 42$ ($t_{s}=38$ for the $\sigma=0.01$ case) for cases B1-B3 are shown in Figure 8.
\begin{figure}
\begin{center}
\includegraphics[width=8.5cm]{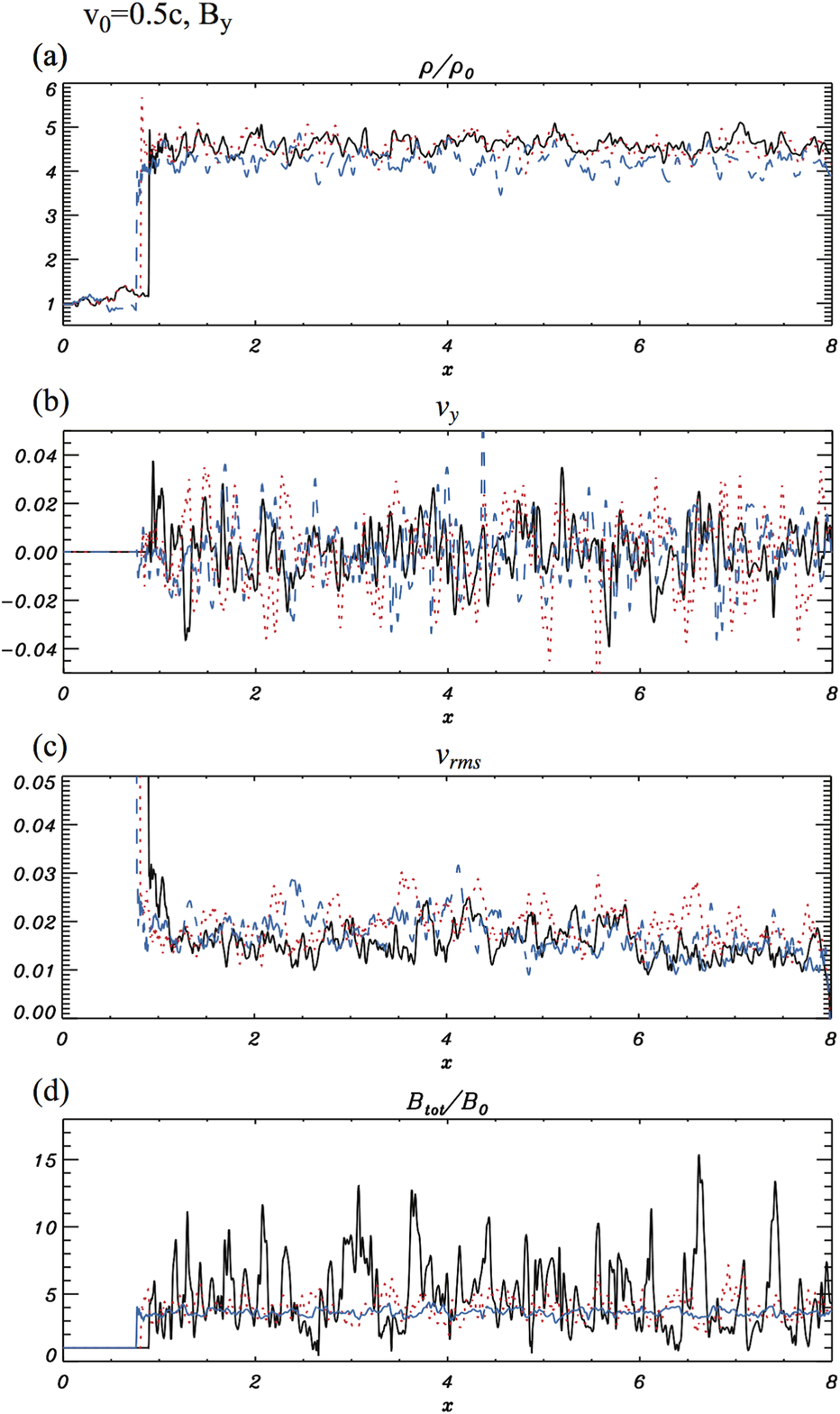}
\end{center}
\caption{The same as in Fig.~3, but for magnetic field perpendicular to the shock propagation direction (cases B1-B3) with $\sigma=0.0001$ (black solid lines), $\sigma=0.001$ (red dotted lines), and $\sigma=0.01$ (blue dashed lines). \label{f8}}
\end{figure}
The shock front is located at $x/L=0.7 - 0.9$. For the perpendicular magnetic field configuration, the shock propagation speed depends slightly on the pre-shock plasma magnetization, and as $\sigma$ becomes larger, the shock velocity increases. Thus for the low and medium $\sigma$ the shock propagation speed is $v_{sh} \simeq 0.17c$ in the contact discontinuity frame, the same as for the parallel initial field case. In the high-$\sigma$ case, the shock velocity is slightly faster, $v_{sh} \simeq 0.18c$. The density jumps by about a factor of 4 in all cases. The transverse velocity profiles show strong velocity fluctuations with a similar root mean square turbulent velocity of $v_{\rm rms} \simeq 0.02c$. This is comparable to the root mean square turbulent velocity obtained for a parallel preshock field. The level of magnetic field shock-compression depends on the magnetization. For low $\sigma$, the magnetic field is amplified by about factor of 4 but the amplification is lower when the magnetization is larger. For high $\sigma$ the amplification factor is only 3.5. The amplitude of the magnetic field fluctuations becomes smaller as $\sigma$ becomes larger.

Figure 9 shows the time evolution of the volume-averaged total magnetic field and the peak total magnetic-field strength in the postshock region for cases B1-B3.
\begin{figure}
\begin{center}
\includegraphics[width=8.5cm]{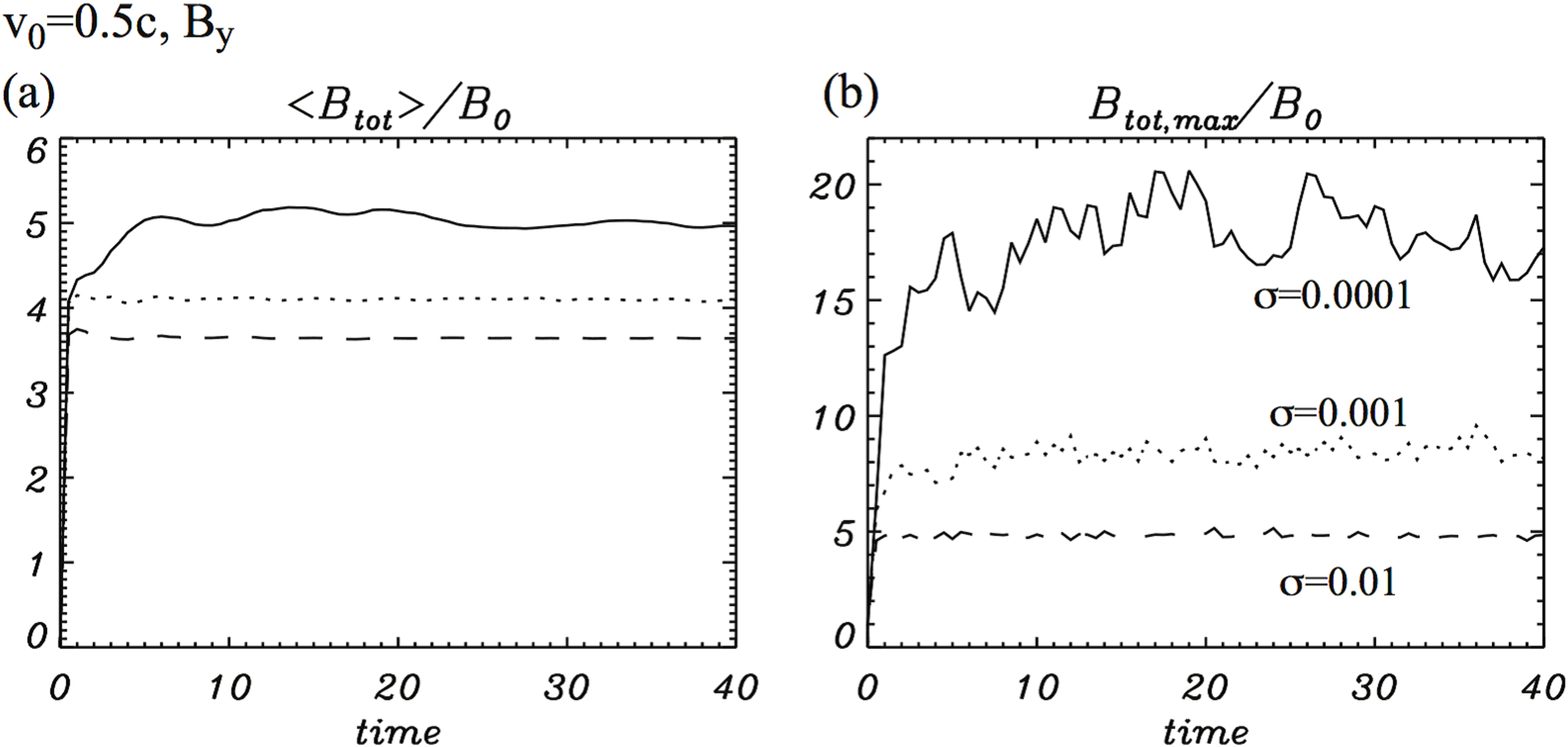}
\end{center}
\caption{The same as in Fig.~4, but for magnetic field perpendicular to the shock propagation direction (cases B1-B3) with $\sigma=0.0001$ (solid lines), $\sigma=0.001$ (dotted lines), and $\sigma=0.01$ (dashed lines).
\label{f9}}
\end{figure}
\begin{figure}
\begin{center}
\includegraphics[width=8.5cm]{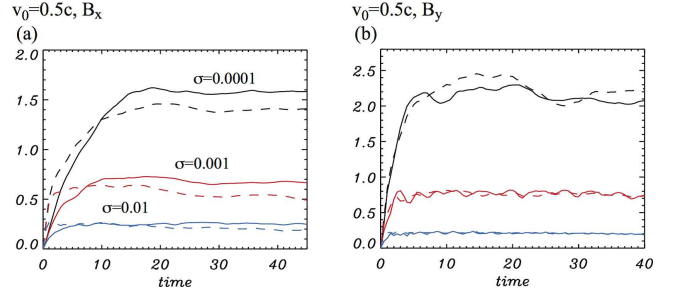}
\end{center}
\caption{Time evolution of the volume-averaged root mean square (rms) fluctuation amplitudes of $\delta B_{x}$ (solid lines) and $\delta B_{y}$ (dashed lines) in the postshock region, normalized to the mean magnetic-field strength in the postshock region. Panel ({\it a}) is for magnetic field parallel, and panel ({\it b}) for magnetic field perpendicular to the shock propagation direction with $\sigma=0.0001$ (black lines), $\sigma=0.001$ (red lines), and $\sigma=0.01$ (blue lines).
\label{f10}}
\end{figure}
The volume-averaged strength of the magnetic field increases at the shock by about a factor of 4 for both the low and medium $\sigma$ cases and by $\sim 3.5$ for the high-$\sigma$ case. The magnetic field is subsequently amplified  by turbulence. In the low-$\sigma$ case, the volume-averaged magnetic field increases up to $\langle B_{\rm tot}\rangle/B_0\sim 5$ and saturates at about $t_s \simeq 6$. This saturation value is larger than in the low-$\sigma$ case with parallel mean magnetic field. 
For the medium $\sigma$ case, magnetic-field amplification by turbulent motion is not significant. The volume-averaged field strength increases to $\langle B_{\rm tot}\rangle/B_0\simeq 4.1$ in the initial shock compression, and maintains this value farther behind the shock. In the high-$\sigma$ case, no turbulent amplification of magnetic field is seen. The behaviour in the medium- and high-$\sigma$ cases (B2-B3) is similar to the behaviour seen in the strong parallel magnetic field cases (A2-A3).

The peak value of the magnetic field is highly variable with time and reaches about $20$ times the initial magnetic-field strength for the low $\sigma$ case B1. For comparison, the peak value of the magnetic field was $15$ in the low-$\sigma$ case A1 with parallel magnetic field. In the medium-$\sigma$ case B2, the peak value is $B_{\rm tot,max}/B_0\sim 10$. This is larger than the maximum of about $6$ times the preshock magnetic field strength seen in the medium-$\sigma$ parallel magnetic field case A2. For the high $\sigma$ case B3, the local maximum value $B_{\rm tot,max}/B_0\sim 5$ is only slightly larger than the volume-averaged strength, again reflecting the fact that magnetic-field amplification is weak in this case (see Fig. 7c).    

In Figure 10 we plot the time evolution of the root mean square fluctuation amplitudes of $\delta B_x$ and $\delta B_y$, normalized to the mean magnetic-field strength in the postshock region (after shock compression for perpendicular cases) to investigate the level of magnetic-field amplification in the turbulent dynamo process for the cases with mildly relativistic flow $v_{0}=0.5c$ and both magnetic field orientations (cases A and B).
In all cases, the fluctuation amplitudes increase and saturate. For parallel shocks (Fig. 10a), $\delta B_y$ initially increases much faster than $\delta B_{x}$, but at saturation $\delta B_x$ is larger than $\delta B_y$.  For perpendicular shocks (Fig. 10b),  the fluctuations $\delta B_x$ and $\delta B_y$ grow similarly and after saturation have about the same strength.
These results indicate that the orientation of the upstream magnetic field relative to the shock normal significantly affects magnetic-field amplification. In the perpendicular case, the magnetic field is enhanced first by shock compression and then by turbulence in the postshock region. The total magnetic-field gain is thus larger in all perpendicular cases than in parallel magnetic field cases. These results are consistent with our earlier study Mizuno et al. (2011b).

\subsection{Dependence on Shock Strength}

Figures 11 and 12 show 1-D cuts along the $x$-axis at $y/L = 0.5$ 
\begin{figure}
\begin{center}
\includegraphics[width=8.5cm]{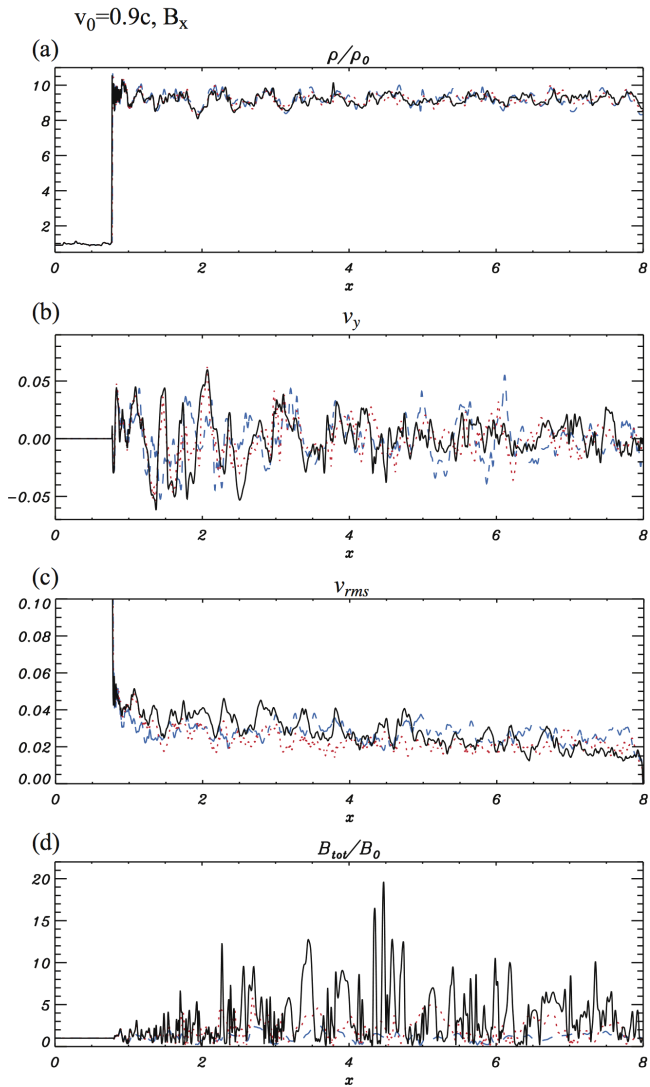}
\end{center}
\caption{The same as in Fig.~3 but for highly relativistic flow (cases C1-C3) at time $t_{s}=24$.
\label{f11}}
\end{figure}
for simulations with parallel magnetic field and different flow speeds.  In particular, Figure 11 shows results for a  highly relativistic flow ($v_{0}=0.9c$) at $t_{s} = 24$ (cases C1-C3), and Figure 12 shows results for a sub-relativistic flow velocity ($v_{0}=0.2c$) at $t_{s}=98$ (cases E1-E3). For highly relativistic inflow, the shock propagation speed in the contact discontinuity frame, $v_{sh} \simeq 0.3c$, is considerably faster than for the mildly relativistic speed, $v_{0}=0.5c$, cases.

For high inflow velocity, $v_{0}=0.9c$,  Eq. \ref{shockspeed} gives $v_{sh}^\prime \simeq 0.945c$. The sound speed in the preshock region is $c_{s}^\prime \simeq 0.04c$ and the relativistic Mach number of the shock is $M_{s}\simeq 70$ with $\gamma_{sh}^\prime\simeq 3$ (see Eq. \ref{mach}).
\begin{figure}
\begin{center}
\includegraphics[width=8.5cm]{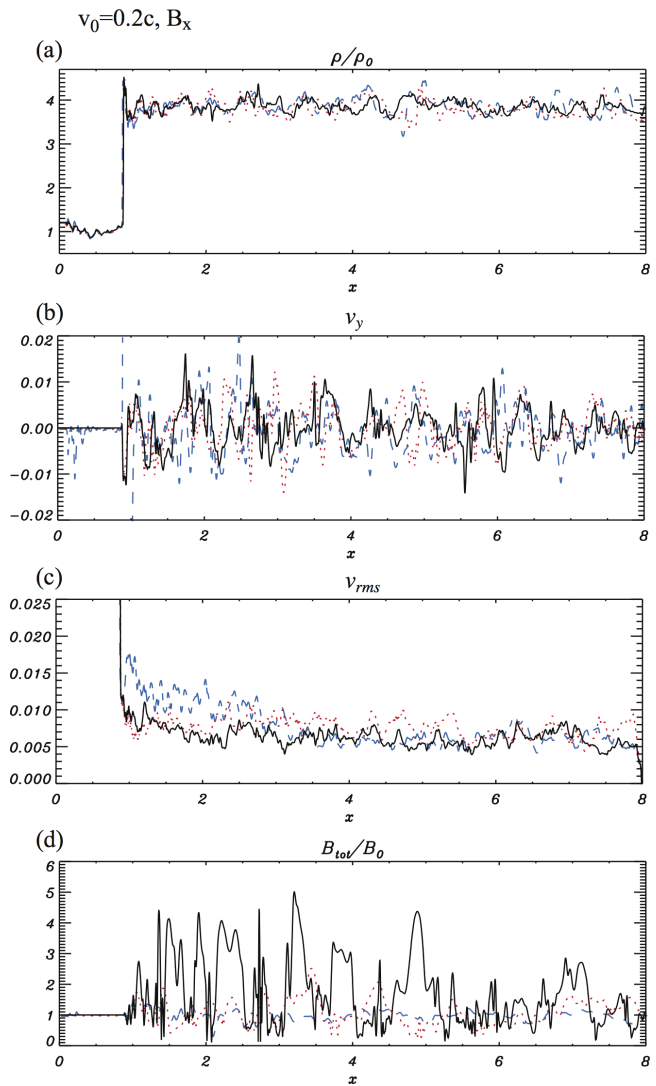}
\end{center}
\caption{The same as in Fig.\ 3 but for sub-relativistic flow (cases E1-E3) at time $t_{s}=98$.}
\label{f12}
\end{figure}
The density jump in the contact discontinuity frame, observed to be a factor 9 in this simulation (Fig. 11b), is close to that expected for a strong relativistic shock (Blandford \& McKee 1976),
\begin{equation}
\frac{n_d}{n_u}=4\,\gamma_0\simeq 9 \ ,
\end{equation}
where $\gamma_0$ is the relative Lorentz factor between the upstream and downstream frames. The continuity condition mandates 
\begin{equation}
v_{sh}^\prime\,\gamma_{sh}^\prime\, n_u= v_{sh}\,\gamma_{sh}\, n_d\ ,
\end{equation}
which is fulfilled for $v_{sh} \simeq 0.3c$.

For cases C1-C3, the transverse velocity fluctuates strongly with $v_{rms} \sim 0.05c$ (Fig. 11b). These velocity fluctuations in the postshock region are sub-relativistic, even though the shock is very strong. The magnetic field is strongly amplified locally when the initial magnetic field is weak (low $\sigma$). In this case the field amplitude achieves more than $15$ times the initial magnetic field strength. Magnetic-field amplification is reduced when the  preshock field is stronger (higher $\sigma$). This trend is similar to that observed for the mildly relativistic inflow cases A1-A3. 

For sub-relativistic flows, the shock propagation speed in the simulation frame is $v_{sh} \simeq 0.07c$, corresponding to $v_{sh}^\prime \simeq 0.27c$, and leads to a shock Mach number $M_{s} \simeq 6.8$. 
The density jumps by only a factor of 3.8, because we are not in the strong-shock limit. In the downstream region, the transverse velocity strongly fluctuates but the maximum is less than $0.02c$. The magnetic field in the postshock region is amplified by a factor of $\sim 5$ above the initial field strength. Magnetic field amplification is weaker than found in the mildly relativistic flow cases. 
\begin{figure}
\begin{center}
\includegraphics[width=8.5cm]{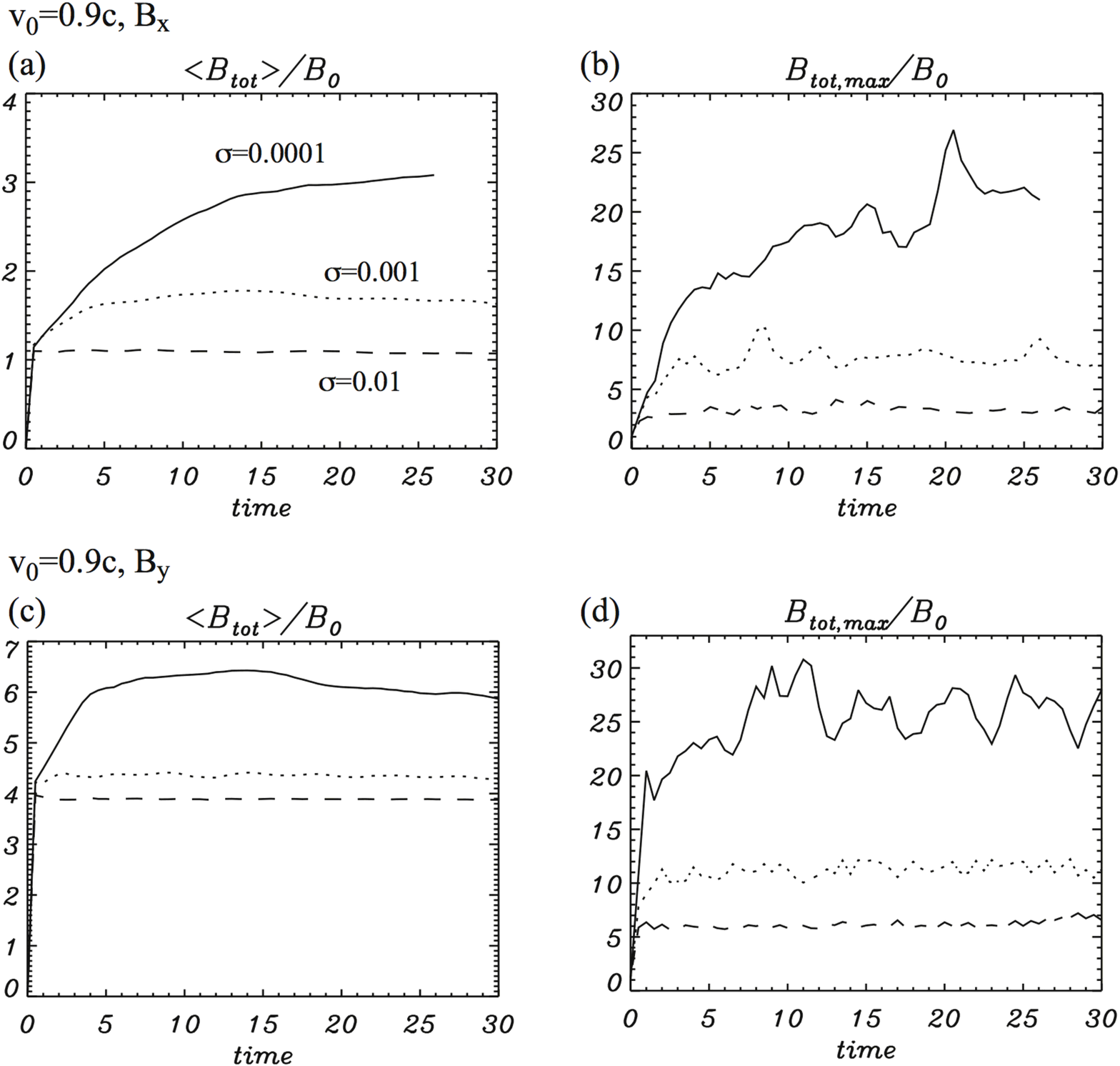}
\end{center}
\caption{Time evolution of (panels a and c) the volume-averaged total magnetic field and (panels b and d) the maximum total magnetic-field strength in the postshock region for a highly relativistic flow velocity, $v_{0}=0.9c$, with parallel (upper) and perpendicular (lower) magnetic field configurations (cases C and D, respectively). Lines indicate different  magnetic field strengths: $\sigma=0.0001$ (solid lines), $\sigma=0.001$ (dotted lines), and $\sigma=0.01$ (dashed lines).
\label{f13}}
\end{figure}

Figure 13 shows the time evolution of the volume-averaged total magnetic field and
the peak total magnetic-field strength in the postshock region for parallel and perpendicular initial magnetic field configurations, with highly relativistic flow, $v_0 = 0.9c$, for the three different initial magnetizations (cases C and D). 

Magnetic-field amplification via turbulent motion happens only when the initial magnetic field is weak. The magnetic energy at saturation is comparable to the turbulent kinetic energy, and magnetic-field amplification in the highly relativistic flow cases is stronger than in the mildly relativistic flow cases (see Figs. 4 and 9) because the stronger shock leads to higher turbulent velocity in the postshock region. For perpendicular shocks with all magnetizations considered, shock compression of the field is more efficient than the turbulent amplification downstream.

\begin{figure}
\begin{center}
\includegraphics[width=8.5cm]{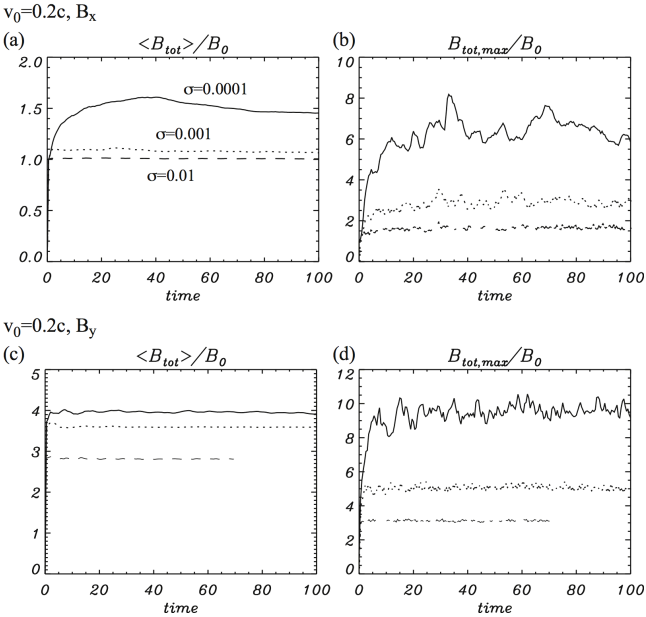}
\end{center}
\caption{Time evolution of ({panels a and c}) the volume-averaged total magnetic field and ({panels b and d}) the maximum total magnetic-field strength in the postshock region for sub-relativistic  flow velocity, $v_{0}=0.2c$, with parallel (upper) and perpendicular (lower) magnetic field configurations (cases E and F, respectively). Lines indicate different  magnetic field strengths: $\sigma=0.0001$ (solid lines), $\sigma=0.001$ (dotted lines), and $\sigma=0.01$ (dashed lines).
\label{f14}}
\end{figure}
For sub-relativistic flows with $v_0 = 0.2c$ (cases E and F), time evolution of the volume-averaged (mean) total magnetic field and the maximum total magnetic-field strength in the postshock region  is shown in Figure 14. 

Weaker turbulence in the postshock region provides less efficient magnetic-field amplification in these cases. For parallel magnetic field cases, the maximum amplification factor, achieved with a low $\sigma$, is at most $\langle B_{\rm tot}\rangle/B_0\sim 1.5$. For perpendicular magnetic field cases we see little amplification beyond the effects of the shock compression, and even shock compression is reduced when $\sigma$ is larger. The local maximum magnetic-field strength reaches about 10 times the initial field strength for the low $\sigma$ case.

Figure 15 shows the time evolution of the root mean square fluctuation amplitudes of $\delta B_x$ and $\delta B_y$, normalized to the mean magnetic field in the postshock region, for three different magnetic-field strengths in the highly relativistic (Fig. 15 a-b) and sub-relativistic (Fig. 15 c-d) flow cases. As for mildly relativistic flows (Fig. 10), the fluctuation amplitude first increases and then saturates in all cases. In parallel shocks, $\delta B_y$ builds up earlier than $\delta B_{x}$, but at the time of saturation, $\delta B_x$ is larger than $\delta B_y$. In perpendicular shocks, the two magnetic-field components grow and saturate similarly. In general, the relative amplitude of field fluctuations increases with shock strength.

\begin{figure}
\begin{center}
\includegraphics[width=8.5cm]{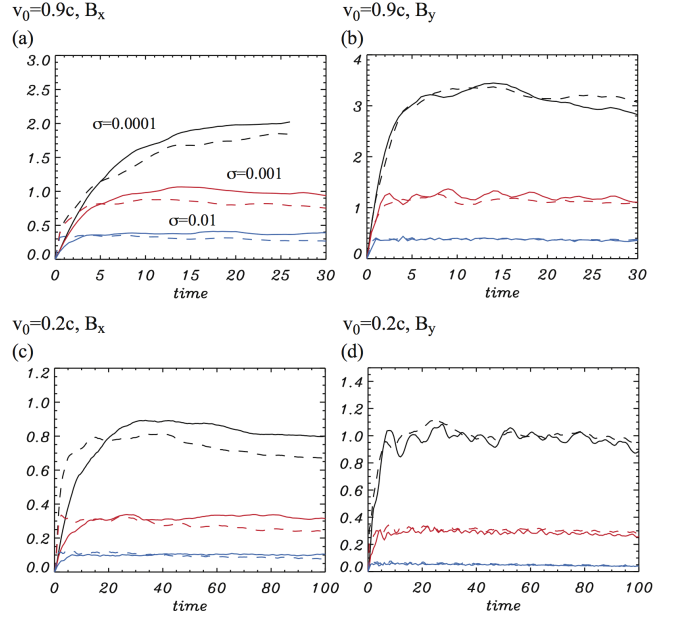}
\end{center}
\caption{Time evolution of the volume-averaged root mean square (rms) fluctuation amplitudes of $\delta B_{x}$ (solid lines) and $\delta B_{y}$ (dashed lines) in the postshock region, normalized to the mean postshock magnetic field, for magnetic field parallel ({\it a, c}) and perpendicular ({\it b, d}) to the shock propagation direction with $\sigma=0.0001$ (black lines), $\sigma=0.001$ (red lines), and $\sigma=0.01$ (blue lines) for highly relativistic flow velocity (upper panels) and  sub-relativistic flow velocity (lower panels).
\label{f15}}
\end{figure}

In summary, as expected our results show that magnetic field amplification strongly depends on the shock strength. A stronger shock leads to a larger density jump, higher turbulent velocity in the postshock region, and a stronger magnetic field at saturation.

\section{Summary and Discussion}

Using a simulation domain of unprecedented length in the flow direction, we have performed two-dimensional relativistic MHD simulations of a relativistic shock propagating through an inhomogeneous medium. Consistent with previous relativistic (Mizuno et al 2011b; Inoue et al. 2011) and non-relativistic studies (Giacalone \& Jokipii 2007; Inoue et al. 2009; Inoue \& Inutsuka 2012; Guo et al. 2012), the postshock magnetic field is amplified through turbulent motions. The amplified magnetic field assumes filamentary structures, and its power spectrum is flatter than Kolmogorov, which is typical for a turbulent dynamo process. 

We find that the saturation level of turbulent field amplification depends on the initial magnetic-field strength. If the initial field is strong, the postshock region becomes turbulent but field amplification does not occur. At a perpendicular shock, the magnetic field is first compressed at the shock and then amplified by turbulent motion in the postshock region. The total field enhancement is larger than at parallel shocks, as observed by Mizuno et al. (2011b).

Generally, saturation occurs when the magnetic energy becomes comparable to the turbulent kinetic energy in the postshock region. This implies that magnetic-field amplification via turbulent motion happens only if the magnetic energy after shock compression is smaller than the turbulent kinetic energy in the postshock region.

In our simulations the turbulent velocity in the postshock region is sub-relativistic and subsonic, even for a strong relativistic shock. All our simulations employ comparable small-amplitude ($\sqrt{\langle\delta \rho^{2}\rangle}/\rho_{0} = 0.012$) density fluctuations in the pre-shock medium. We note that analytical work by Sironi \& Goodman (2007) (see also Goodman \& MacFadyen 2008) indicates that the energy density of vortical motions, generated in an interaction between an ultra-relativistic shock and a small-amplitude density fluctuation, increases with the square of the density contrast (i.e., the postshock turbulent velocity grows linearly with the density fluctuation amplitude).
Thus, larger density perturbations lead to greater deformations of the shock front and produce stronger vorticity in the postshock region. As a result, the magnetic field can be more efficiently amplified. This analytical prediction has been recently confirmed in MHD relativistic shock simulations by Inoue et al. (2011) (see also Guo et al. 2012 for the nonrelativistic case), who showed that the rate of the initial exponential growth of postshock magnetic turbulence is proportional to the amplitude of the velocity fluctuations. They also found that the shock-induced velocity dispersion can approach the postshock sound speed if the shock propagates in a medium with large-amplitude density inhomogeneities. However, 
it is not clear whether supersonic relativistic turbulence can be produced and maintained downstream of the shock.  For example, in simulations initialized with turbulence by Inoue et al. (2011),
the kinetic energy in relativistic turbulence decayed much faster than kinetic energy in transonic turbulence due to dissipation of relativistic turbulence into internal energy via shocklets. It remains to be verified that the same result applies to the case of turbulence driven by ultra-relativistic shock propagation through an upstream medium containing strong density contrasts.

In this paper we have performed simulations in two-dimensional geometry to take long simulation region in shock propagation direction in order to follow the saturation of magnetic-field amplification by turbulence in postshock region which did not achieved previous paper (Mizuno et al. 2011b). In general the turbulence structure, for example the slope of the Kolmogorov-like power spectrum, is different in two and three dimensions. Although we expect that the possible difference is not significant (see Inoue et al. 2009), to more realistically analyze three-dimensional phenomena we will extend the current investigation to three-dimensional simulations in future work. 

In GRBs, afterglow modeling suggest that the ratio of the magnetic energy density to the internal energy density, $\epsilon_B$, has a broad range of values in the emission region. Our simulation results imply that the broad range suggested by the observations is indicative of different properties in the circumburst medium, e.g., different amplitude of density fluctuations, and magnetic-field strength and direction, notwithstanding  differences in shock strength. The strong variability in the prompt emission phase requires maintenance of strong relativistic turbulence in the emission region. If prompt emission were due to jet-medium interaction, then this would require a weak medium magnetic field strength. More likely, GRB prompt emission is of an internal origin, which may invoke collisions of highly magnetized shells (Zhang \& Yan 2011). Our simulations in this paper do not apply to this regime, and future work is needed to test whether substantial reconnection-driven turbulence can be generated and maintained in a magnetically dominated flow.

In AGN blazars, multi-waveband monitoring often finds rapidly variable gamma-ray flares that typically accompany lesser variability and/or increase in the emission at other wavebands. 
``Multi-wavelength light curves of gamma-ray bright blazars reveal strong correlations across wavebands, 
yet striking dissimilarities in the details. The linear polarization tends to be highly variable in both 
degree and position angle, which implies that the magnetic field is turbulent" (Marscher 2013). In general, the emission characteristics based on a turbulent plasma crossing a standing conical shock associated with the millimetre-wave core in VLBA images of blazars agrees with the characteristics of multi-waveband light curves and polarization variations (Marscher 2011). In particular, multi-waveband monitoring of AO 0235+164 and OJ 287 indicates gamma-ray flares associated with enhanced emission from the 43 GHz radio core that is consistent with this turbulent cell model for variability (Marscher \& Jorstad 2010; Agudo et al. 2011b, 2011c). In these sources the radio core is located at considerable distance (parsecs) from the central engine, AO 0235+164 (Agudo et al. 2012) and OJ 287 (Agudo et al., 2011a). 
Our simulation results suggest that strong density inhomogeneities in the preshock relativistic jet fluid would lead to the development of strong turbulence in the post recollimation shock region from which the observed emission originates. The highest photon energies would come from the smallest regions containing the highest magnetic fields and exhibit the most rapid time variability.  Structure in the preshock magnetic field would influence the ordering of the postshock turbulent field depending on the magnetization of the preshock jet fluid

\section*{Acknowledgments}
This work has been supported by NSF awards AST-0908010, and AST-0908040 to UA and UAH, AST-0908362 to UNLV and NASA awards NNX08AG83G and NNX12AH06G to UAH. Y.M. acknowledges support from Taiwan National Science Council award NSC 100-2112-M-007-022-MY3. The work of J.N. has been supported by the Polish National Science Centre through projects DEC-2011/01/B/ST9/03183 and DEC-2012/04/A/ST9/00083.
M.P. acknowledges support by the Helmholtz Alliance for Astroparticle Physics, HAP, funded by the Initiative and Networking Fund of the Helmholtz Association. The simulations were performed on the Columbia and Pleiades Supercomputer at the NAS Division of the NASA Ames Research Center, the SR16000 at YITP in Kyoto University, and the Nautilus at the National Institute for Computational Sciences in the XSEDE project supported by National Science Foundation.

\bsp

\label{lastpage}


\begin{thebibliography}{}

\bibitem[Agudo et al. (2011a)]{Agu11a} Agudo I. et al.,  2011a, ApJ, 726, L13

\bibitem[Agudo et al. (2011b)]{Agu11b} Agudo I. et al.,  2011b, in Proc. 2011 Fermi Symposium, eConf C110509, arXiv:1110.6463

\bibitem[Agudo et al. (2011c)]{Agu11c} Agudo I. et al.,  2011c, ApJet al., 735, L10

\bibitem[Agudo et al. (2012)]{Agu12} Agudo I. et al.,  2012, Int. J. Mod. Phys. Conf. Ser., 08, 271

\bibitem[Aharonian et al. (2003)]{Aha03} Aharonian F. et al.,  2003, A\&A, 410, 813

\bibitem[Aharonian et al. (2007)]{2007ApJ...664L..71A} Aharonian F. et al.,  2007, ApJ, 664, L71 

\bibitem[Albert et al.(2007)]{Alb07} Albert J. et al.,  2007, ApJ, 669, 862

\bibitem[Arlen et al.(2013)]{2013ApJ...762...92A} Arlen T. et al.,  2013, ApJ, 762, 92 

\bibitem[Bamba et al.(2003)]{Bam03} Bamba A., Yamazaki R., Ueno M., \& Koyama K., 2003, ApJ, 589, 827

\bibitem[Bamba et al.(2005a)]{Bam05a} Bamba A., Yamazaki R., Yoshida T., Terasawa T., \& Koyama K., 2005a, ApJ, 621, 793

\bibitem[Bamba et al.(2005b)]{Bam05b} Bamba A., Yamazaki R., \& Hiraga J.S., 2005b, ApJ, 632, 294

\bibitem[Barniol Duran(2013)]{Bar13} Barniol Duran R., 2013, arXiv: 1311.1216

\bibitem[Begelman et al.(2008)]{Beg08} Begelman M.C., Blandford R.D., \& Rees M.J., 2008, MNRAS, 384, L19

\bibitem[Bell(2004)]{bell04} Bell A.R., 2004, MNRAS, 353, 550

\bibitem[Blandford \& McKee(1976)]{1976PhFl...19.1130B} Blandford R.D., \& McKee C.F., 1976, Physics of Fluids, 19, 1130 

\bibitem[Brandenburg \& Subramanian(2005)]{Bra05} Brandenburg A. \& Subramanian K., 2005, Phys. Rep., 417, 1

\bibitem[Brouillette(2002)]{Bro02} Brouillette M., 2002, Annu. Rev. Fluid Mech., 34, 445

\bibitem[Bykov et al.(2008)]{2008ApJ...689L.133B} Bykov A.M., Uvarov Y.A., \& Ellison D.C., 2008, ApJ, 689, L133 

\bibitem[Caprioli \& Spitkovsky(2013)]{caprioli13} Caprioli D., \& Spitkovsky A., 2013, ApJ, 765, L20

\bibitem[Cho \& Lazarian(2003)]{Cho03} Cho J. \& Lazarian A., 2003, MNRAS, 345, 325

\bibitem[Cho et al.(2009)]{Cho09} Cho J., Vishniac E.T., Beresnyak A., Lazarian A., \& Ryu D., 2009, ApJ, 693, 1449

\bibitem[Del Zanna et al.(2007)]{Del07} Del Zanna L., Zanotti O., Bucciantini N., \& Londrillo P., 2007, A\&A, 473, 11

\bibitem[Fraschetti(2013)]{Fra13} Fraschetti F. 2013,, ApJ, 770, 84

\bibitem[Ghisellini \& Tavecchio(2008)]{Ghi08} Ghisellini G., \& Tavecchio F., 2008, MNRAS, 386, L28

\bibitem[Giacalone \& Jokipii(1999)]{Giac99} Giacalone J., \& Jokipii J.R., 1999, ApJ, 520, 204

\bibitem[Giacalone \& Jokipii(2007)]{Giac07} Giacalone J., \& Jokipii,J.R., 2007, ApJ, 663, L41

\bibitem[Giannios et al.(2009)]{Gia09} Giannios D., Uzdensky D.A., \& Begelman M.C., 2009, ApJ, 395, L29

\bibitem[Goodman \& MacFadyen(2008)]{Goo08} Goodman J. \& MacFadyen A., 2008, J. Fluid. Mech., 604, 325

\bibitem[Gruzinov(2001)]{Gru01} Gruzinov A., 2001, ApJ, 563, L15

\bibitem[Guo et al.(2012)]{Guo12} Guo F., Li S., Giacalone J., Jokipii J.R., \& Li D., 2012, ApJ, 747, 98

\bibitem[Inoue(2012)]{Ino12b} Inoue T., 2012, ApJ, 760, 43

\bibitem[Inoue \& Inutsuka(2012)]{Ino12a} Inoue T., \& Inutsuka S., 2012, ApJ, 759, 35

\bibitem[Inoue et al.(2009)]{Ino09} Inoue T., Yamazaki R., \& Inutsuka S., 2009, ApJ, 695, 825

\bibitem[Inoue et al.(2011)]{Ino11} Inoue T., Asano K., \& Ioka K., 2011, ApJ, 731, 77

\bibitem[Inoue et al.(2012)]{Ino12} Inoue T., Yamazaki R., Inutsuka S., \& Fukui Y., 2012, ApJ, 744, 71

\bibitem[Kato \& Takabe(2008)]{kato08} Kato T.N., \& Takabe H., 2008, ApJ, 681, L93

\bibitem[Kazantsev(1968)]{Kaz68} Kazantsev A.P., 1968, Sov. Phys.-JTEP Lett., 26, 1031

\bibitem[Komissariv(1997)]{Kom97} Komissarov S.S., 1997, Phys. Lett. A, 232, 435

\bibitem[Krawczynski et al.(2004)]{Kra03} Krawczynski H. et al., 2004, ApJ, 601, 151

\bibitem[Lazar et al.(2009)]{Laz09} Lazar A., Nakar E., \& Piran T., 2009, ApJ, 695, L10

\bibitem[Levinson(2007)]{Lev07} Levinson A., 2007, ApJ, 671, L29

\bibitem[Marscher(2011)]{Mar11} Marscher A.P., 2011, in Proc. Fermi Symposium, eConf C110509, arXiv:1110.6463

\bibitem[Marscher(2013)]{Mar13} Marscher A.P., 2013, AAS, 221, 339.53

\bibitem[Marscher \& Jorstad(2010)]{Mar10a} Marscher A.P., \& Jorstad S.G., 2010, in Proc. Fermi meets Jansky - AGN in Radio and Gamma-ray, ed. T. Savolainen et al. 171, arXiv:1005.5551

\bibitem[Marscher et al.(1992)]{Mar92} Marscher A.P., Gear W.K., \& Travis J.P., 1992, in Blazar Variability, ed. E. Valtaoja \& M. Valtonen (Cambridge Univ. Press), 85

\bibitem[Marscher et al.(2010)]{Mar10b} Marscher A.P., et al.,  2010, ApJ, 630, L5

\bibitem[Medvedev \& Loeb(1999)]{Med99} Medvedev M.V. ,\& Loeb A., 1999, ApJ, 526, 697

\bibitem[M\'{e}sz\'{a}ros(2006)]{Mes06} M\'{e}sz\'{a}ros P., 2006, Rep. Prog. Phys. 69, 2259

\bibitem[Mignone et al.(2005)]{Mig05} Mignone A., Plewa T., \& Bodo G., 2005, ApJS, 160, 199

\bibitem[Milosavljevi\'c \& Nakar(2006)]{mil06} Milosavljevi\'c M., \& Nakar E., 2006, ApJ, 651, 979

\bibitem[Mizuno et al.(2006)]{Miz06} Mizuno Y., Nishikawa K.-I., Koide S., Hardee P., \& Fishman G.J., 2006, arXiv astro-ph, 0609004

\bibitem[Mizuno et al.(2011a]{Miz11a} Mizuno Y., Hardee P.E., \& Nishikawa K.-I., 2011a, ApJ, 734, 19

\bibitem[Mizuno et al.(2011b)]{Miz11b} Mizuno Y., Pohl M., Niemiec J., Zhang B., Nishikawa K.-I., \& Hardee P.E. 2011b, ApJ, 726, 62

\bibitem[Narayan \& Kumar(2009)]{Nar09} Narayan R., \& Kumar P., 2009, MNRAS, 394, L117

\bibitem[Niemiec et al.(2008)]{niem08} Niemiec J.,  Pohl M., Stroman T., \& Nishikawa K.-I., 2008, ApJ, 684, 1174

\bibitem[Niemiec et al.(2010)]{niem10} Niemiec J.,  Pohl M., Bret A., \& Stroman T., 2010, ApJ, 709, 1148

\bibitem[Niemiec et al.(2012)]{niemiec12} Niemiec J.,  Pohl M., Bret A., \& Wieland V., 2012, ApJ, 759, 73

\bibitem[Nishikawa et al.(2005)]{Nis05} Nishikawa K.-I., Hardee P., Richardson G., Preece R., Sol H., \& Fishman G.J., 2005, ApJ, 622, 927.

\bibitem[Nishikawa et al.(2009)]{Nis09} Nishikawa K.-I., et al., 2009, ApJ, 698, L10

\bibitem[Palma et al.(2008)]{Pal08} Palma G., Mignone A., Vietri M., \& Del Zanna L., 2008, ApJ, 686, 1103

\bibitem[Panaitescu(2005)]{Pan05} Panaitescu A., 2005, MNRAS, 363, 1409

\bibitem[Panaitescu \& Kumar(2002)]{Pan02} Panaitescu A., \& Kumar P., 2002, ApJ, 571, 779

\bibitem[Parker(1971)]{Par71} Parker E.N., 1971, ApJ, 163, 255

\bibitem[Piran(2005)]{Pir05} Piran T., 2005, Rev. Mod. Phys., 76, 114

\bibitem[Pohl et al.(2005)]{pyl05} Pohl M., Yan H., \& Lazarian A., 2005, ApJ 626, L101

\bibitem[Ramirez-Ruiz et al.(2005)]{Ram05} Ramirez-Ruiz E., Garc\'{i}a-Segura G., Salmonson J.D., \& P\'{e}rez-Rend\'{o}n B., 2005, ApJ, 631, 435

\bibitem[Riquelme \& Spitkovsky(2010)]{2010ApJ...717.1054R} Riquelme M.A., \& Spitkovsky A., 2010, ApJ, 717, 1054 

\bibitem[Riquelme \& Spitkovsky(2009)]{2009ApJ...694..626R} Riquelme M.A., \& Spitkovsky A., 2009, ApJ, 694, 626 

\bibitem[Sano et al.(2012)]{San12} Sano T., Nishihara K., Matsuoka C., \& Inoue T., 2012, ApJ, 758, 126

\bibitem[Santana, Barniol Duran, \& Kumar(2013)]{Sant13} Santana R., Barniol Duran R., \& Kumar P., 2013, arXiv:1309.3277

\bibitem[Schekochihin \& Cowley(2007)]{Sch07} Schekochihin A., \& Cowley S., 2007, in Magnetohydrodynamics - Historical Evolution and Trends, ed S. Molokov, R. Moreau, \& H. Moffatt (Berlin: Springer) 85

\bibitem[Schekochihin et al.(2004)]{Sch04} Schekochihin A.A., Cowley S.C., Taylor S.F., Maron J.L., \& McWilliams J.C., 2004, ApJ, 612, 276

\bibitem[Sironi \& Goodman(2007)]{Sir07} Sironi L., \& Goodman J., 2007, ApJ, 671, 1858

\bibitem[Spitkovsky(2008)]{Spi08} Spitkovsky A., 2008, ApJ, 682, L5

\bibitem[Stroman et al.(2009)]{2009ApJ...706...38S} Stroman T., Pohl M., \& Niemiec J., 2009, ApJ, 706, 38 

\bibitem[Synge(1957)]{Syn57} Synge J.L., 1957, The Relativistic Gas (Amsterdam: North-Holland)

\bibitem[Taub(1948)]{Tau48} Taub A.H., 1948, Phys. Rev., 74, 328

\bibitem[Uchiyama et al.(2007)]{Uch07} Uchiyama Y., et al., 2007, Nature, 449, 576

\bibitem[Vink \& Laming(2003)]{Vin03} Vink J., \& Laming J.M., 2003, ApJ, 584, 758

\bibitem[Yost et al.(2003)]{Yos03} Yost S.A., Harrison F.A., Sari R., \& Frail D.A., 2003, ApJ, 597, 459

\bibitem[Zhang et al.(2009)]{WZha09} Zhang W., MacFadyen A., \& Wang P., 2009, ApJ, 690, L40

\bibitem[Zhang \& Yan(2010)]{BZha11} Zhang B., \& Yan H., 2011, ApJ, 726, 90

\end{thebibliography}
\end{document}